\documentclass[aip,jmp,amsmath,amssymb,preprint]{revtex4-2}
\usepackage{amssymb,amsthm}
\usepackage{graphicx}
\graphicspath{{./}{Figures/}{figures/}}
\usepackage{subcaption}
\usepackage{array}

\newtheorem{theorem}{Theorem}
\newtheorem{proposition}{Proposition}

\theoremstyle{remark}
\newtheorem{remark}{Remark}

\DeclareMathOperator{\meas}{meas}
\newcommand{\Z}{\mathbb{Z}}

\newcommand{\Tt}{\mathbb{T}}

\begin{document}

\title{Spectral Regimes of Truncated sinc-Power and Pulse-Train Kernels:
Edge Exponents, Per-Edge Plunges, and the Landau--Widom Limit}

\author{Unnikrishna Pillai}
\affiliation{Department of Electrical and Computer Engineering,
New York University Tandon School of Engineering, Brooklyn, NY 11201, USA}

\date{\today}

\begin{abstract}
A signal band-limited to one rectangular band and observed through a finite window is the classical concentration problem of Slepian, Landau, Pollak, and Widom. The eigenvalues of the operator cluster near one and near zero, separated by a transition called the plunge. The plunge width grows like the logarithm of the window length, with a universal coefficient of one over pi squared set by the band-edge jump. This paper follows the clusters and the plunge as the band is deformed. An elementwise power of the kernel smooths the band edge, and the plunge gives way to a power-law edge with exponent one less than the power. Disjoint copies of the band widen the plunge in proportion to the number of band edges, as the multiple-interval theorem of Landau and Widom implies, so the coefficient is a per-edge constant. Shrinking each band to a line gives the classical finite-rank operator. Overlapping bands give a multilevel staircase, and a self-similar band gives a cluster hierarchy governed by the Cantor dimension. At the smooth edge, rescaled eigenvalues do not collapse onto one curve. The edge descriptor passes from a constant to an exponent to an integer to a fractal dimension. For the single band, classical results place all eigenfilter zeros on the unit circle, with more of them inside the band as the order rises, so the dominant eigenvectors act as lowpass filters.
\end{abstract}


\maketitle

\section{Introduction}
A signal whose spectrum fills a single rectangular band, seen through a finite time window, is
the oldest problem in spectral concentration. Let the passband have fractional half-width $W$
and let the window hold $N$ samples. The finite-window autocorrelation of the flat band is the
sinc sequence $a_k=\sin(2\pi Wk)/(\pi k)$, and the concentration operator is the Hermitian
Toeplitz matrix $T_N(a)$ built from it. Its eigenvectors are the discrete prolate spheroidal
sequences (DPSS), and each eigenvalue measures how much of one mode the window retains
\cite{slepian61,landaupollak61,landaupollak62,slepian78}.

The eigenvalues lie in $(0,1)$ and split sharply into two groups. About $2WN$ lie near $1$ and
the rest near $0$, where $2W$ is the fractional bandwidth and $N$ the window length. A thin
transition band, the \emph{plunge}, lies between them. Landau and Widom \cite{landauwidom80}
fixed its width, and Slepian \cite{slepian78} gave the discrete form: for a threshold $\alpha\in(0,1)$,
\begin{equation}
\#\{n:\lambda_n>\alpha\}=2WN+\frac{1}{\pi^2}\log\!\frac{1-\alpha}{\alpha}\,\log N+o(\log N),
\label{eq:lw}
\end{equation}
so the symmetric plunge count is $\frac{2}{\pi^2}\log\frac{1-\alpha}{\alpha}\log N$. The constant
$1/\pi^2$ is universal, independent of the bandwidth and the threshold. This is the classical
Slepian--Landau--Pollak--Widom picture, and it is the reference point for the rest of the paper.

The plunge belongs to the symbol, not to the kernel. The matrix $T_N(a)$ is Toeplitz with
generating symbol $f(\omega)$, the frequency-domain shape, here the indicator of the band.
Equation \eqref{eq:lw} is the eigenvalue-counting form of the two-term trace asymptotics for a
discontinuous symbol, and the logarithmic term with its constant is carried by the jump of $f$
at the band edge \cite{widom82,sobolev13,gioevklich06}.

The rectangular band is a single box in frequency. This paper asks what becomes of the two
clusters and the plunge when the box is deformed. Each deformation acts directly on the
symbol.

\emph{(i) Hadamard powers.} Replacing the DPSS matrix by its elementwise $p$-th power
multiplies the generating sequence pointwise and hence convolves the symbol
$p$ times. The box becomes an order-$p$ B-spline: continuous for $p\ge 2$, with the band
edge softened from a jump to an algebraic zero.

\emph{(ii) Frequency pulse trains.} Replicating the pass band into $M$ disjoint copies
keeps the symbol a $\{0,1\}$ indicator but multiplies the number of band edges.

\emph{(iii) Atomic limit.} Shrinking each band to zero width turns the symbol into a sum
of Dirac masses; the kernel becomes a pure cosine sum.

Beyond these three, the same symbol calculus governs several further cases treated below:
overlapping pulses and multilevel staircases, the self-similar (devil's-staircase) limit
with its Cantor dimension, the raised-cosine pulse-shaping family, and a worked ``hat''
symbol in which a smooth cap rides a discontinuous pedestal.

\noindent\textbf{Contributions.}
\begin{itemize}
\item[C1.] A power-law spectral edge for sinc-power kernels: at a fixed fraction below the
support edge the eigenvalues fall as the power $p-1$ of the distance to the edge, independent of
the bandwidth (Theorem~\ref{thm:edge}). The law is the quantile form of Szeg\H{o}'s theorem
\cite{kms53} applied to the exact edge monomial of the B-spline symbol, and it is tested
numerically for $p=2,\dots,6$; the plunge of \eqref{eq:lw} is recovered as the singular endpoint
$p\to1$.
\item[C2.] A per-edge reading of the multiband plunge: $M$ disjoint bands give a plunge
$M$ times that of one band, the discrete form of the multiple-interval theorem of Landau and
Widom \cite{landauwidom80}, so $1/\pi^2$ is a per-edge constant (Proposition~\ref{prop:comb}).
\item[C3.] The classical finite-rank limit of atomic symbols \cite{grenanderszego,pisarenko73}:
a sum of $M$ cosines yields rank $2M$, with closed-form nonzero eigenvalues for one cosine and no
plunge (Proposition~\ref{prop:cos}).
\item[C4.] A self-similar (devil's-staircase) symbol yields a pure-point eigenvalue cluster
hierarchy whose count-versus-mass exponent is the Cantor dimension $D=\log2/\log3$
(Proposition~\ref{prop:devil}).
\item[C5.] A unifying statement: the regimes are ordered by symbol regularity, and the descriptor
of the spectral edge changes from a constant to an exponent to an integer, and to a fractal
dimension in the self-similar limit, with Landau--Widom the shared limiting case.
\item[C6.] On the eigenvector side, two classical facts combine for the single band: Slepian's
zero count \cite{slepian78} puts $k$ zeros of the $k$-th eigenfilter inside the band, and Trench's
theorem \cite{trench94} places all eigenfilter zeros on the unit circle. Proposition~\ref{prop:zeros}
records the conjugate-pair bookkeeping, $\lfloor k/2\rfloor$ pairs, and the resulting lowpass
character of the dominant eigenvectors.
\end{itemize}

\section{Setup: Operators and Symbols}
\label{sec:setup}
The development uses the time--frequency duality familiar to engineers \cite{papoulispillai}: the
kernel lives in the time domain and its symbol (spectrum) in the frequency domain. The
prototype is $\sin(2\pi Wt)/(\pi t)$ in time, whose transform is an even rectangular
pulse of unit height $h=1$ on the angular frequency band $|\omega|<2\pi W$; the discrete kernel
below is its sampled counterpart. Throughout, ``pulse'' refers to such a rectangular block in the
frequency domain, ``height'' to its amplitude $h$ ($h=1$ for a single pulse, with
overlapping pulses stacking so the symbol takes the heights $0,h,2h,\dots$), and
``half-width'' to the fractional half-width $W$ (angular half-width $2\pi W$); a \emph{pulse
train} is a set of equal pulses of half-width $W$ with centers spaced by $2B$ in the same
fractional units, disjoint for $B\ge W$ and overlapping for $B<W$. All spectral
structure is read off from this frequency-domain symbol.

For a sequence $\{r_k\}_{k\in\Z}$ with $r_{-k}=\overline{r_k}$ let $T_N(r)$ denote the $N\times N$
Hermitian Toeplitz matrix $[T_N(r)]_{mn}=r_{m-n}$, $0\le m,n\le N-1$. Its \emph{symbol} is the
real function $f$ on $\Tt=[-\pi,\pi)$, or a measure for line spectra, with Fourier coefficients
$r_k=\frac{1}{2\pi}\int_{\Tt}f(\omega)\,e^{ik\omega}\,d\omega$, so that $f(\omega)=\sum_k r_k e^{-i\omega k}$;
$T_N(f)$ denotes the same matrix, and an even symbol gives a real symmetric matrix. The DPSS
matrix is $A=T_N(a)$ with
\begin{equation}
a_k=\frac{\sin(2\pi W k)}{\pi k},\quad a_0=2W,\qquad
f_1(\omega)=\mathbf{1}_{\{|\omega|<2\pi W\}},
\label{eq:box}
\end{equation}
the indicator of the pass band, of fractional measure $2W$. Non-asymptotic and finite-$N$ bounds on the eigenvalues of $T_N(a)$ are given in \cite{osipov13,davenport21,bonami21}.

The Hadamard (Schur) product is the key device. For Toeplitz matrices the elementwise
product is again Toeplitz, with the generating sequences multiplied pointwise; and by the
convolution theorem, multiplying two sequences pointwise in the time domain convolves their
symbols in the frequency domain. Raising the DPSS matrix to its $p$-th Hadamard power
therefore convolves the box symbol with itself $p$ times:
\begin{equation}
A^{\circ p}=T_N(a^p),\qquad f_p=\underbrace{f_1*\cdots*f_1}_{p},
\label{eq:bspline}
\end{equation}
the order-$p$ B-spline \cite{schoenberg73,deboor01} supported on $|\omega|<2\pi pW$ (assuming $pW<\tfrac12$ to avoid
aliasing). Here $*$ denotes normalized circular convolution on the torus,
$(u*v)(\omega)=\frac{1}{2\pi}\int_{-\pi}^{\pi}u(\omega-\theta)\,v(\theta)\,d\theta$, the same normalization used in
Appendix~\ref{app:bspline}. Each self-convolution is a smoothing pass (convolving with a box is a moving
average), so successive Hadamard powers round the symbol off progressively: $f_p$ is a
piecewise polynomial of degree $p-1$ and smoothness class $C^{p-2}$. A single elementwise
operation thus moves the symbol from a discontinuous indicator ($p=1$, the box, with a jump
at the band edge) to a triangle ($p=2$, continuous with a corner) and on to ever smoother
splines, the band edge softening from a jump into an algebraic zero of order $p-1$. The support fraction of $f_p$ is written $s_p:=2pW$.

Two further constructions act on the symbol just as transparently. Replicating the band to
centers $\{\xi_m\}$ gives the kernel $a_k\sum_m\cos(\xi_m k)$, a single box times a
Dirichlet factor, whose symbol is the corresponding union of shifted boxes. Letting the
band width tend to zero sends each box to a Dirac mass; for the symbol
$\pi\sum_j[\delta(\omega-\omega_j)+\delta(\omega+\omega_j)]$ the kernel is the cosine
sum $\sum_j\cos(\omega_j k)$.

The eigenvalue distribution of $T_N(f)$ for a bounded real symbol is governed by the
Szeg\H{o} distribution theorem \cite{grenanderszego,bottchersilbermann,gray06}, which is
the single tool used throughout this paper, so it is worth stating in operational terms.
It says that for large $N$ the histogram of the eigenvalues of $T_N(f)$ converges to the
histogram of the values taken by the symbol $f(\omega)$ as $\omega$ ranges uniformly over
$\Tt$. Formally, the empirical spectral measure converges weakly to the push-forward of
normalized Lebesgue measure under $f$; informally, an eigenvalue near a level $\lambda$ is
as common as the frequency band on which $f$ sits near $\lambda$. Sorting the eigenvalues
in descending order therefore traces out the symbol's level sets \cite{kms53}: the eigenvalue at index
$n$ converges to the upper quantile $Q(n/N)$ of $f$, the level exceeded on exactly a
fraction $n/N$ of the band. Two consequences drive everything below. Where $f$ is flat
(constant on a set of positive measure) the eigenvalues pile up into a \emph{cluster} at
that height; where $f$ jumps discontinuously the sorted eigenvalues would have to jump too,
and the finite matrix can only approximate the jump by a narrow ramp, the \emph{plunge}.
The regularity of the symbol thus dictates the spectrum, and the three regimes below are
simply the three ways a symbol can be regular: smooth, piecewise constant, or atomic.

\section{Smooth Symbols and the Edge Exponent}
\label{sec:smooth}
For $p\ge2$ the symbol $f_p$ is continuous, so by Szeg\H{o} the eigenvalues fill the whole
interval $[0,\max f_p]$; there is no two-point clustering and, as confirmed below, no
$\log N$ plunge. Smoothing the band edge has removed the jump that forced the eigenvalues
into two clusters, so they now spread continuously between the noise floor and the peak.
The informative quantity is then no longer a plunge width but the \emph{rate} at which the
smallest eigenvalues approach zero. These are the marginally concentrated modes, and their
decay sets how gracefully the usable signal dimension fades. That rate is the spectral-edge
exponent.

\begin{theorem}[Edge-exponent law]
\label{thm:edge}
Let $\hat f_p=f_p/f_p(0)$ be the peak-normalized order-$p$ B-spline symbol, where
$f_p(0)=\sum_k a_k^p$ is the symbol peak (asymptotically equal to the largest eigenvalue
$\lambda_{\max}$), with support fraction $s_p=2pW$, $pW<\tfrac12$, $p\ge2$. For every fixed
fraction $u$ with $s_p-4W\le u<s_p$, the descending-ordered eigenvalues $\lambda_n$ of $T_N(a^p)$
satisfy
\begin{equation}
\lim_{N\to\infty}\frac{\lambda_{\lfloor uN\rfloor}}{f_p(0)}=C_p\,\big[\pi\,(s_p-u)\big]^{\,p-1},
\qquad C_p=\frac{1}{(p-1)!\,(2\pi)^{p-1}f_p(0)},
\label{eq:edge}
\end{equation}
where the exponent $p-1$ is independent of $W$ and of normalization. This is a quantile
(macroscopic) edge law: it holds at a fixed fraction below the support edge, not at a fixed
index from the edge.
\end{theorem}

\begin{proof}
By the quantile form of the Szeg\H{o} theorem \cite{kms53} the descending eigenvalue
$\lambda_{\lfloor uN\rfloor}$ converges, at
continuity points of the limiting quantile, to the level $\lambda$ for which the upper-level fraction
$u(\lambda)=\frac{1}{2\pi}\meas\{\omega:\hat f_p(\omega)>\lambda\}$ equals $u$. A $p$-fold
self-convolution of a box is a piecewise polynomial of degree $p-1$ whose outermost piece
is an \emph{exact} monomial: near each support endpoint $\omega_e=2\pi pW$, writing
$\epsilon=\omega_e-\omega$,
\begin{equation}
\hat f_p(\omega)=C_p\,\epsilon^{\,p-1},\qquad 0\le\epsilon\le 4\pi W,
\label{eq:endpoint}
\end{equation}
with $C_p$ as in \eqref{eq:edge}, derived in Appendix~\ref{app:bspline}. For a small level $\lambda$ the symbol
exceeds $\lambda$ everywhere on its support except a shell of width
$\delta=(\lambda/C_p)^{1/(p-1)}$ at each of the two edges, whence
\begin{equation}
u(\lambda)=s_p-\frac{2\delta}{2\pi}=s_p-\frac1\pi\Big(\frac{\lambda}{C_p}\Big)^{1/(p-1)} .
\end{equation}
Setting $u(\lambda)=u$ and solving for $\lambda$ gives \eqref{eq:edge}. Because
\eqref{eq:endpoint} is exact, not merely leading order, on the whole outermost spline piece,
the limit holds for every $u$ in the stated range.
\end{proof}

The result is the quantile form of the Szeg\H{o} distribution theorem \cite{kms53} applied to the
exact edge monomial of the B-spline. It fixes the eigenvalue at a fixed fraction below the support
edge and is distinct from the classical extreme-eigenvalue laws for a symbol with a zero of
finite order \cite{parter61,widom61,bottchergrudsky98}, which fix the smallest eigenvalue at a
fixed index; here the symbol vanishes on a whole interval. Toeplitz matrices built from cardinal
B-splines also arise in isogeometric analysis \cite{garoni14}, where the spline values are the
matrix entries rather than the symbol.

\begin{remark}[The Landau--Widom limit]
As $p\downarrow1$ the exponent $p-1\to0$, so the right side of \eqref{eq:edge} tends to a
constant: the algebraic edge stiffens into a jump. At $p=1$ the symbol is the box and the
finite-$N$ rounding of that jump is the $\frac{1}{\pi^2}\log N$ plunge of
\eqref{eq:lw}. The smooth edge law and the Landau--Widom plunge are thus two faces of one
family indexed by the edge regularity $p-1$.
\end{remark}

\begin{figure}[t]
\centering
\begin{subfigure}{0.3\linewidth}\includegraphics[width=\linewidth]{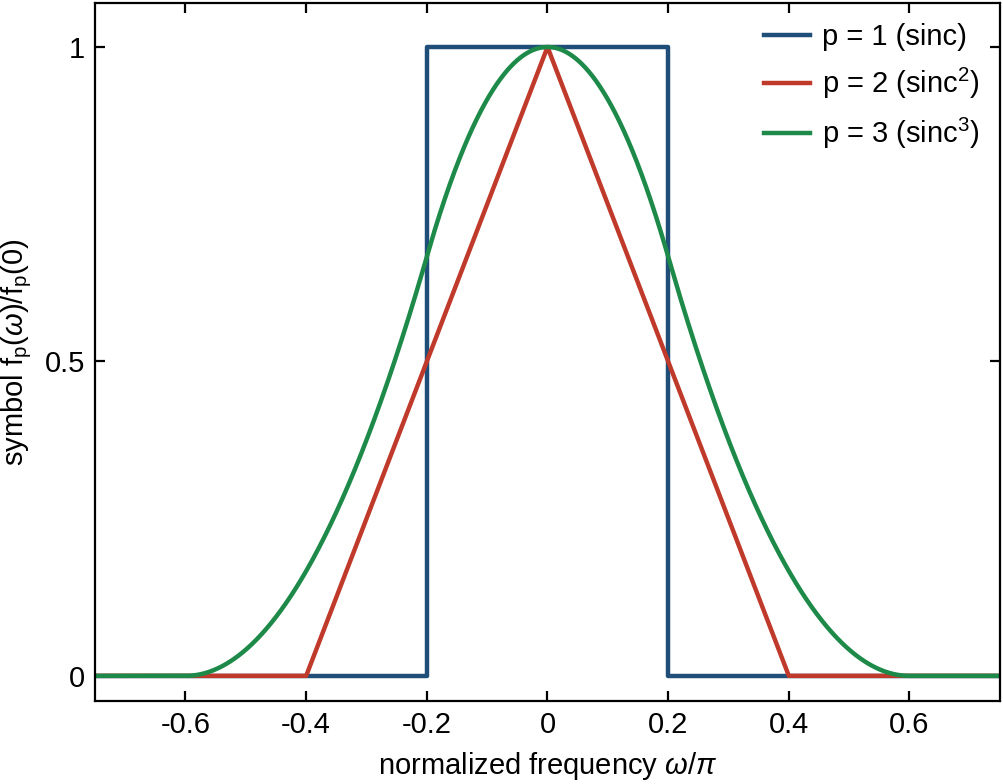}\caption{}\end{subfigure}\hfil
\begin{subfigure}{0.3\linewidth}\includegraphics[width=\linewidth]{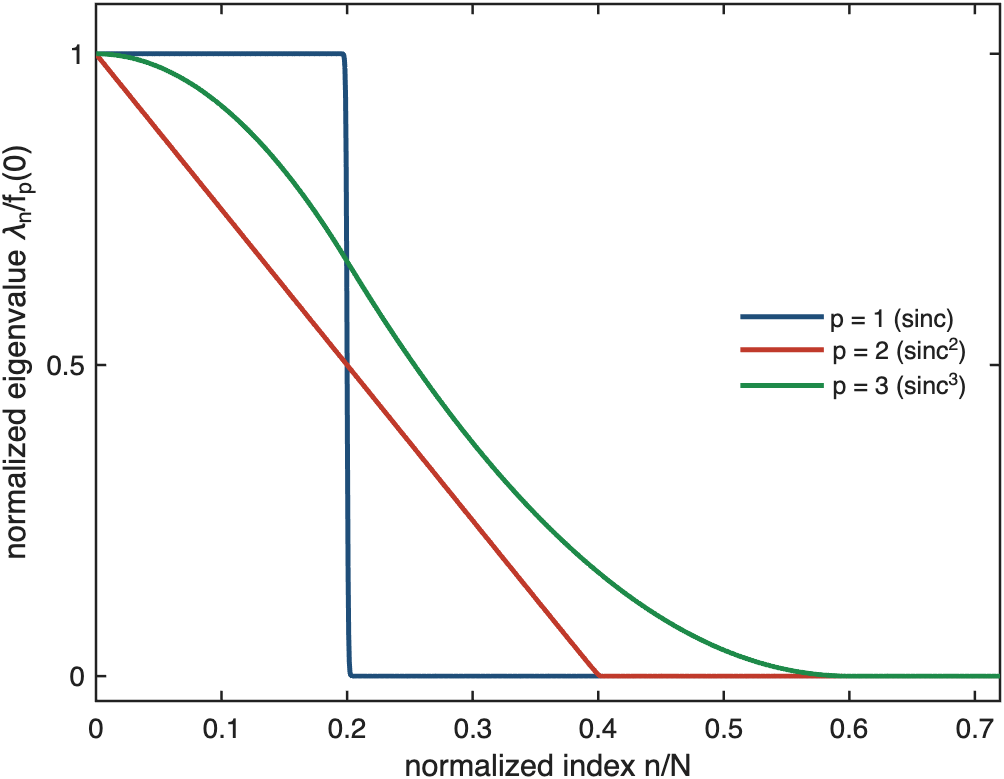}\caption{}\end{subfigure}\hfil
\begin{subfigure}{0.3\linewidth}\includegraphics[width=\linewidth]{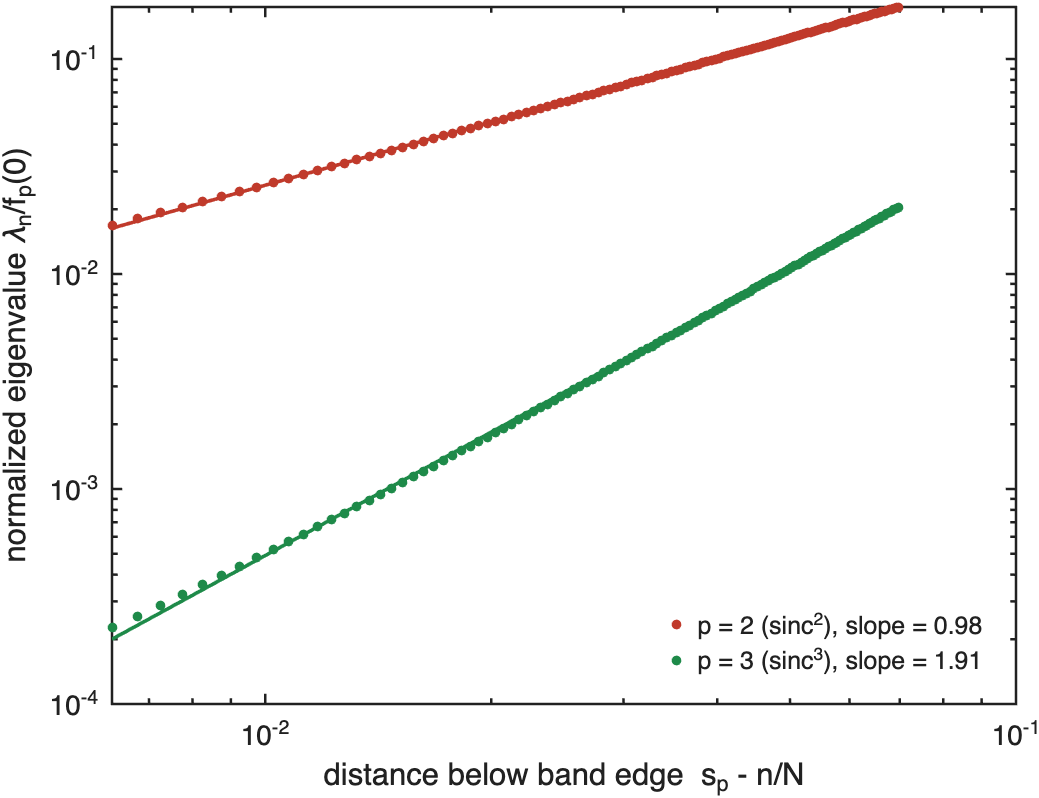}\caption{}\end{subfigure}
\caption{Smooth regime, $W=0.1$, $N=2000$. (a) Peak-normalized symbols $\hat f_p$: box
($p{=}1$, a jump), triangle ($p{=}2$, a corner), quadratic B-spline ($p{=}3$, $C^1$).
(b) Descending-ordered normalized eigenvalues versus $n/N$: a sharp plunge at $2W$ for
$p{=}1$, an exact linear ramp ending at $s_2=0.4$ for $p{=}2$, and a smooth fill ending at
$s_3=0.6$ for $p{=}3$. (c) Band-edge log--log fit of \eqref{eq:edge}: fitted exponents
$0.98$ ($p{=}2$) and $1.91$ ($p{=}3$) against the theoretical $p-1$.}
\label{fig:smooth}
\end{figure}

Fig.~\ref{fig:smooth} illustrates the mechanism. Fig.~\ref{fig:smooth}(a) shows the symbols: a box
($p=1$) with its jump, a triangle ($p=2$) with a corner, and a smooth $C^1$ spline
($p=3$). Fig.~\ref{fig:smooth}(b) shows the resulting spectra: the $p=1$ step at $n/N=2W$, the exact
linear ramp for $p=2$ (the quantile of a triangle is linear), and a smooth fill for $p=3$.
Fig.~\ref{fig:smooth}(c) checks \eqref{eq:edge} on log--log axes; the fitted slopes, given in
the caption, agree with $p-1$.

\begin{figure}[t]
\centering
\includegraphics[width=0.6\linewidth]{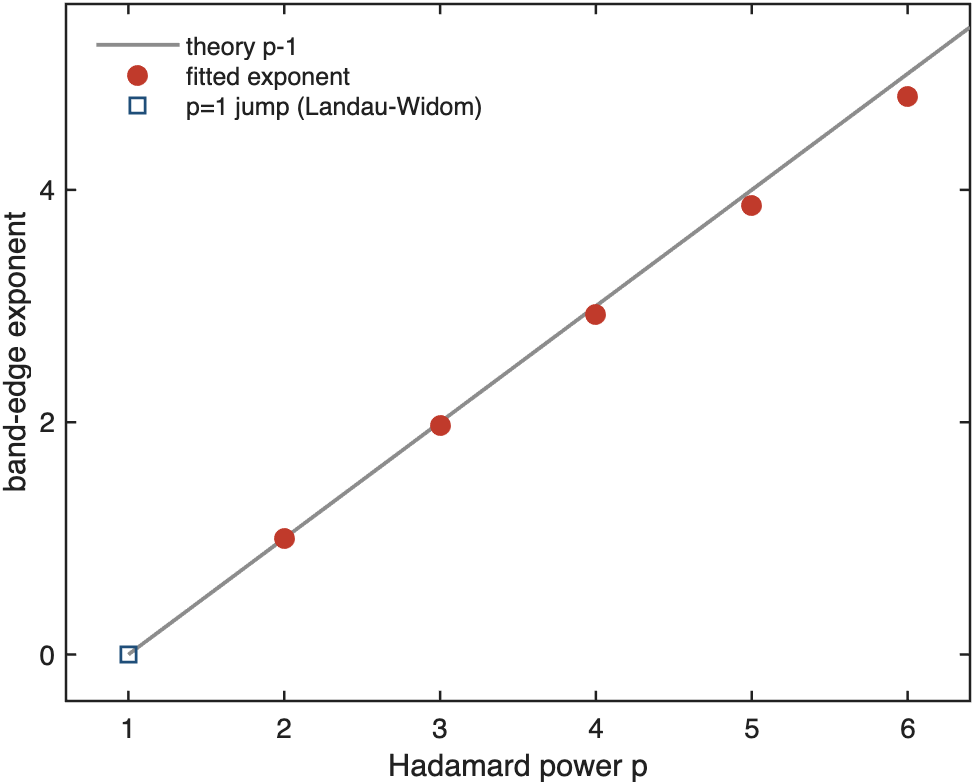}
\caption{Fitted band-edge exponent versus Hadamard power $p$ (filled circles) against the
exact law $p-1$ (line). The open square at $p=1$ marks the jump (exponent $0$), the
Landau--Widom endpoint.}
\label{fig:sweep}
\end{figure}

Fig.~\ref{fig:sweep} plots the fitted band-edge exponent against the power $p$ for $p=2,\dots,6$, together with the exact law $p-1$ and the $p=1$ jump as the exponent-zero endpoint. The fitted exponent tracks $p-1$ across the sweep, the small residual growing with $p$ because a steeper edge is more sensitive to the finite-$N$ rounding at the inner end of the fit window. The full spectrum matches the Szeg\H{o} quantile over the same range.

\subsection{Pulse shaping: the raised-cosine family}
\label{sec:rc}
\begin{figure}[t]
\centering
\begin{subfigure}{0.49\linewidth}\includegraphics[width=\linewidth]{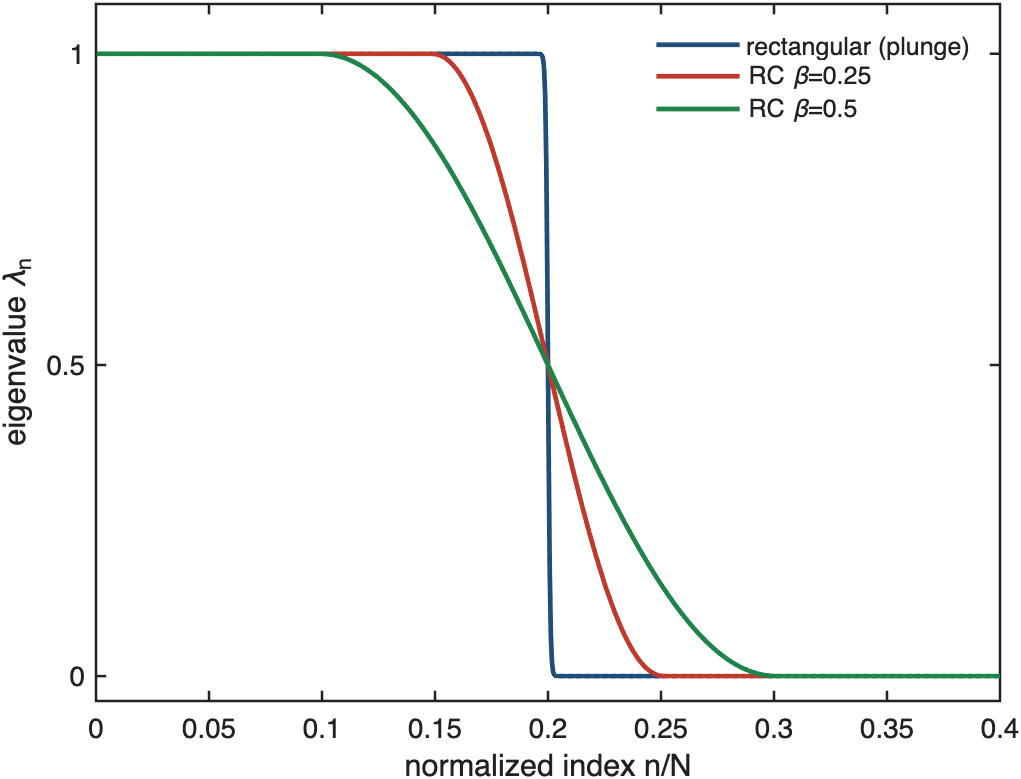}\caption{}\end{subfigure}\hfil
\begin{subfigure}{0.48\linewidth}\includegraphics[width=\linewidth]{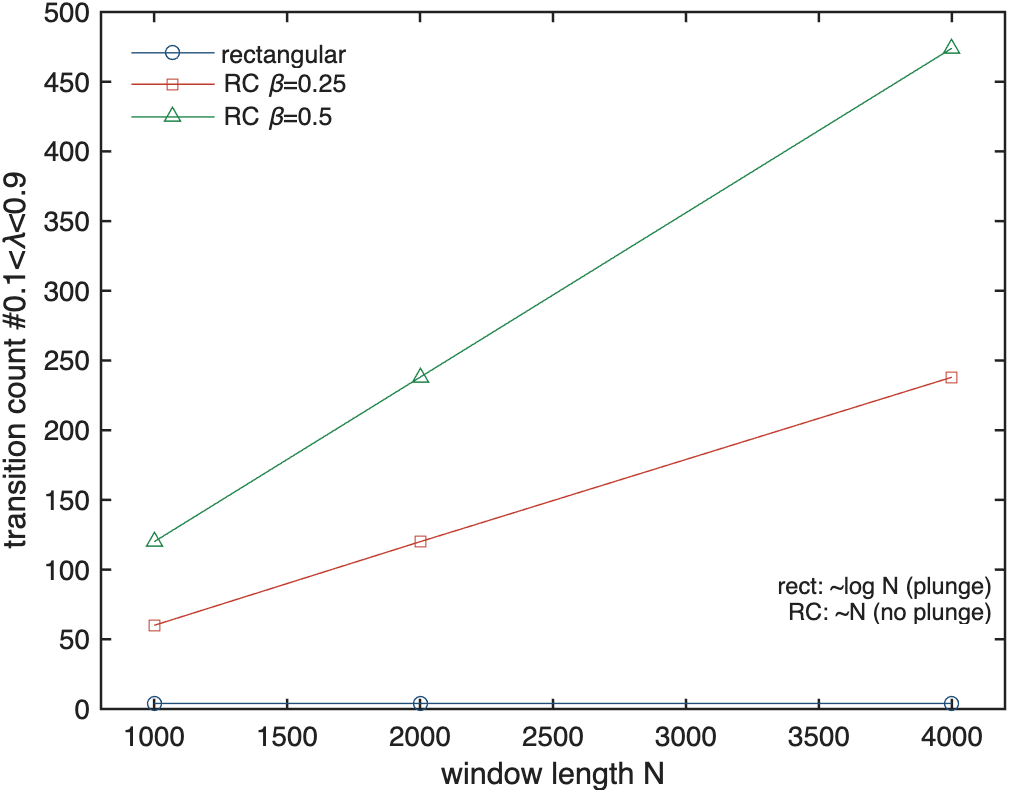}\caption{}\end{subfigure}
\caption{Raised-cosine pulse shaping, $\omega_0=0.2\pi$, $N=2000$. (a) Descending eigenvalue
spectra: the rectangle plunges vertically at $n/N=\omega_0/\pi$, while the raised cosine
($\beta=0.25,0.5$) descends smoothly over the roll-off band. (b) Transition count
$\#\{0.1<\lambda<0.9\}$ versus $N$: the rectangle stays flat ($\sim\log N$, a vanishing
fraction, the plunge), while the raised cosine grows linearly ($\sim N$, a fixed
fraction, no plunge). Across $N=1000,2000,4000$ the box count stays near $4$, while the
raised-cosine ($\beta=\tfrac12$) count grows $120\to238\to474$.}
\label{fig:rc}
\end{figure}

The smooth regime is the natural home of communications pulse shaping, where the abrupt
band edge of the ideal (sinc) filter is deliberately tapered to suppress sidelobes and
intersymbol interference. The standard choice is the raised-cosine filter \cite{proakis}
\begin{equation}
H_\beta(\omega)=
\begin{cases}
1, & |\omega|\le\omega_-,\\
\tfrac12\big[1+\cos(\pi x)\big], & \omega_-<|\omega|\le\omega_+,\\
0, & |\omega|>\omega_+,
\end{cases}
\label{eq:rc}
\end{equation}
with roll-off $\beta\in[0,1]$, half-amplitude bandwidth $\omega_0$, band edges
$\omega_\mp=\omega_0(1\mp\beta)$, and normalized roll-off coordinate
$x=(|\omega|-\omega_-)/(\omega_+-\omega_-)\in[0,1]$. Taking $H_\beta$ as the symbol places
the kernel squarely in the smooth regime: $H_\beta$ is continuous (indeed $C^1$, with zero slope at
both ends of the roll-off), so it has no jump and therefore no plunge. The spectrum instead
shows a cluster at $1$ from the flat top, of fractional size $\omega_0(1-\beta)/\pi$,
followed by a smooth descent to zero across the roll-off band, as in
Fig.~\ref{fig:rc}(a).

The contrast with the rectangle is sharpest in how the transition scales with $N$. For the
box the transition set $\{n:\alpha<\lambda_n<1-\alpha\}$ holds only $O(\log N)$
eigenvalues, a vanishing fraction (the Landau--Widom plunge), whereas for the raised cosine
it holds a \emph{fixed} fraction, proportional to $\beta$, of all $N$ eigenvalues.
The box transition count barely moves with $N$, while the raised-cosine ($\beta=\tfrac12$)
count grows in proportion to $N$ (Fig.~\ref{fig:rc}(b)). A fixed fraction is a distributed
transition, not a plunge.

The roll-off also fixes the edge exponent through the local behavior of $H_\beta$ where it
leaves zero. The cosine taper meets zero quadratically, so by the same quantile-edge argument used in
Theorem~\ref{thm:edge} the
raised-cosine band edge carries exponent $2$; the root-raised
cosine $\sqrt{H_\beta}$, used when the shaping is split between transmitter and receiver,
leaves zero linearly and carries exponent $1$. Finite-$N$ fits reproduce both exponents. The
pulse-shaping family thus
traces the same exponent ladder as the Hadamard powers of Theorem~\ref{thm:edge}: the
rectangle is the jump (exponent $0$, the plunge), the root-raised cosine is linear (exponent
$1$), and the raised cosine is quadratic (exponent $2$). As $\beta\downarrow0$ the roll-off
band shrinks to nothing, $H_\beta$ stiffens back into the box, and the plunge re-forms, so
$\beta$ continuously interpolates between plunge and no plunge, which is precisely the
engineering trade-off between a maximally sharp set of concentrated signaling dimensions and
a graceful, low-sidelobe roll-off.

\begin{remark}[Smooth but not band-limited]
What removes the plunge is smoothness of the symbol, not band-limitation. A damped cosine
$r_k=e^{-\gamma|k|}\cos(\theta_0k)$ has symbol $\tfrac12[P_\gamma(\omega-\theta_0)+P_\gamma(\omega+\theta_0)]$, a pair of Poisson
kernels (discrete Lorentzians) with $P_\gamma(\theta)=\sinh\gamma/(\cosh\gamma-\cos\theta)$. This symbol
is smooth and strictly positive, with no jump and no support edge, so it too has no
plunge. Its eigenvalues form a smoothly decaying profile, and the transition count again
grows like $N$. As $\gamma\downarrow0$ the Lorentzians sharpen toward Dirac masses and the kernel
approaches the rank-$2$ cosine of the atomic regime; for every $\gamma>0$ it is a smooth, plunge-free
interpolant between the smooth and atomic regimes.
\end{remark}

\section{Indicator Symbols and Per-Edge Plunges}
\label{sec:indicator}
A union of disjoint pass bands is still a $\{0,1\}$ indicator, hence still (asymptotically)
a projection: the spectrum retains the two clusters and the plunge. What changes is the
plunge \emph{width}. For time and frequency sets that are finite unions of intervals, Landau
and Widom \cite{landauwidom80} showed that the logarithmic term grows with the number of
intervals, and the trace asymptotics tie it to the number of jumps in the symbol
\cite{widom82,sobolev13}. The discrete multiband form, in per-edge terms, is the following.

\begin{proposition}[Per-edge multiband plunge]
\label{prop:comb}
Let the symbol be the indicator of $M$ pairwise-disjoint bands, each of fractional width
$2W$, with the gaps between bands bounded below by a fixed positive distance as $N\to\infty$, so
the symbol has $2M$ jump discontinuities. Then $\lambda_{\max}\to1$, the upper
cluster has fractional size $2WM$, and the symmetric plunge count obeys
\begin{equation}
\#\{n:\alpha<\lambda_n<1-\alpha\}\sim\frac{2M}{\pi^2}\,\log\!\frac{1-\alpha}{\alpha}\,\log N .
\label{eq:multiplunge}
\end{equation}
\end{proposition}

\begin{proof}
Write the pass band as $M$ disjoint intervals $I_j$ of equal length and set
$Q_j=T_N(\mathbf 1_{I_j})$, so $0\le Q_j\le I$ and $T_N(\mathbf 1_\Omega)=\sum_j Q_j$
with spectrum in $[0,1]$. Modulation by $D_j=\operatorname{diag}(e^{-i\xi_j n})$ carries
each band to the baseband prolate, $D_jQ_jD_j^{*}=Q_0$, so every $Q_j$ has the single-band
spectrum: a cluster of $2WN$ eigenvalues near $1$ and the transition of \eqref{eq:lw}
\cite{slepian78,widom82}. For $i\ne j$ the interval indicators are disjoint, hence
$\sum_{k\in\mathbb Z}q^{(i)}_k\overline{q^{(j)}_k}=0$ and
$\operatorname{Tr}(Q_iQ_j)=-\sum_{|k|<N}|k|\,q^{(i)}_k\overline{q^{(j)}_k}$, which stays bounded in
$N$. The summand is of order $|k|^{-1}$, so the sum does not converge absolutely; it is bounded
through the oscillatory phase $e^{i(\xi_i-\xi_j)k}$ with $\xi_i\ne\xi_j$, by an Abel--Dirichlet
estimate, and the same bound holds for $\|Q_iQ_j\|_1$. In a smooth
transition trace $\operatorname{Tr}\varphi(\sum_jQ_j)$ with $\varphi(0)=\varphi(1)=0$ the mixed
terms are then $O(1)$, so the transition traces add and the plunge is $M$ times that of one
band, giving \eqref{eq:multiplunge}; this is the discrete form of the multiple-interval theorem
of Landau and Widom \cite{landauwidom80} (see also \cite{sobolev13}).
\end{proof}

The intuition is that each band edge acts as an independent source of spectral ambiguity.
A finite window of length $N$ cannot resolve a discontinuity in the symbol: near each jump
the sorted eigenvalues must bridge the gap from $1$ to $0$, and they do so over a band of
width $\propto\log N$ whose coefficient is the Landau--Widom constant. Edges that are far
apart smear disjoint sets of eigenvalues, so their contributions simply add. A single band
has two edges and gives the plunge \eqref{eq:lw}; $M$ disjoint bands have $2M$ edges and
give $M$ times that plunge. Formally this is the multiple-interval theorem of Landau and Widom
\cite{landauwidom80}, the one-dimensional case of the Wiener--Hopf trace asymptotics
\cite{sobolev13}: each edge contributes an independent Fisher--Hartwig logarithmic term and
$1/\pi^2$ is the per-edge coefficient. The message is compact: the Landau--Widom constant is
not a property of the band but of its edges: each discontinuity contributes $1/\pi^2$,
independently of the others. A regular pulse train places the centers at $\xi_m=m\Delta$ in angular frequency, for
which the kernel factors as $a_k\,D_M(\Delta k)$ with $D_M$ the Dirichlet kernel; since each band
has angular half-width $2\pi W$, disjointness requires $\Delta>4\pi W$ and the outermost band edge
below $\pi$. The practical reading is that
fragmenting a fixed total bandwidth into more sub-bands does not change the size of the
top cluster ($2WM$ stays fixed if $WM$ does) but inflates the ambiguous transition in
direct proportion to the number of edges.

\begin{figure}[t]
\centering
\begin{subfigure}{0.46\linewidth}\includegraphics[width=\linewidth]{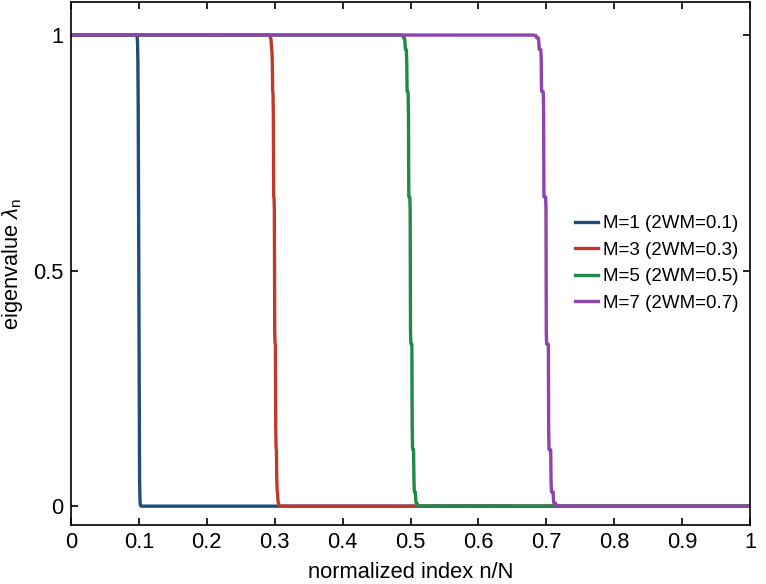}\caption{}\end{subfigure}\hfil
\begin{subfigure}{0.46\linewidth}\includegraphics[width=\linewidth]{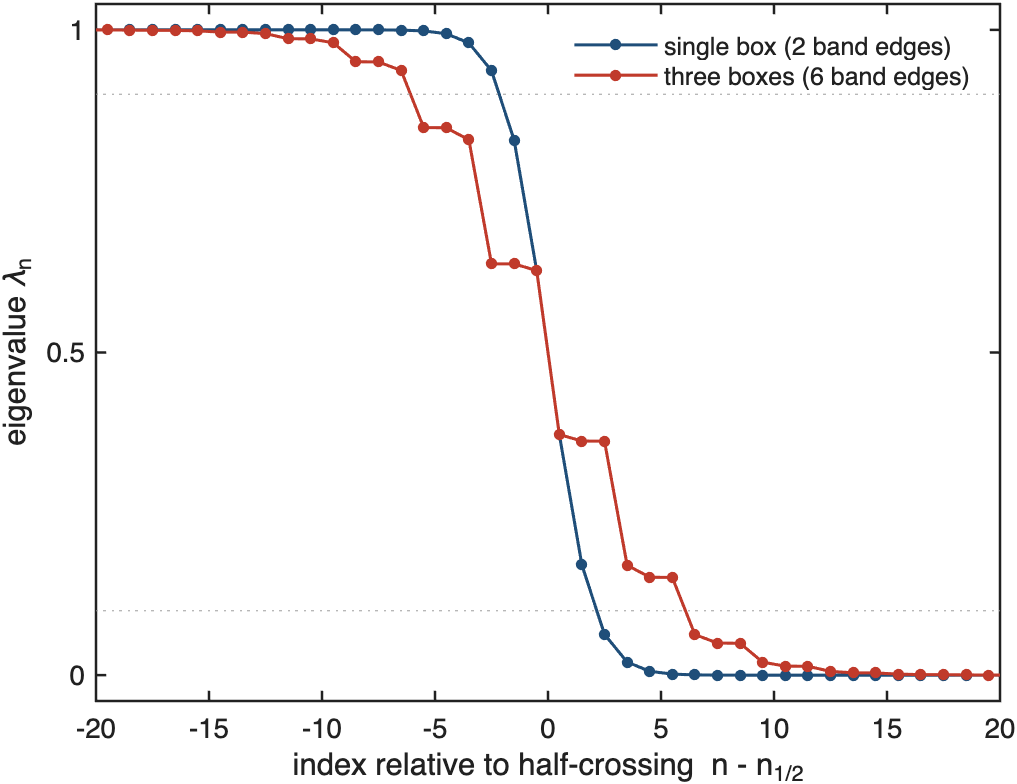}\caption{}\end{subfigure}
\caption{Indicator regime, $W=0.05$ (a) and $W=0.08$ (b), $N=2000$. (a) Descending spectra
for a frequency pulse train of $M=1,3,5,7$ disjoint bands; the upper cluster ends at
$2WM$, and the plunge widens with $M$. (b) Plunge profiles aligned at the half-crossing
$n_{1/2}$: the three-band plunge ($6$ edges) is asymptotically three times as wide as the
single-band plunge ($2$ edges); dotted lines mark $\alpha=0.1,\,1-\alpha=0.9$.}
\label{fig:indicator}
\end{figure}

Fig.~\ref{fig:indicator} confirms \eqref{eq:multiplunge}. Fig.~\ref{fig:indicator}(a) shows the spectra of
$M=1,3,5,7$ equal disjoint bands; the upper cluster ends at $2WM$ in each case.
Fig.~\ref{fig:indicator}(b) aligns the single-band and three-band plunges at their half-crossings: the
three-band transition is three times as wide, edge for edge, and one can see incipient
sub-steps from the six individual edges. Numerically the plunge count grows in proportion to
the number of band edges: across $M=1,2,3,5,7$ the measured count tracks $M$ times the
single-band plunge, approaching the exact ratio $M$ of \eqref{eq:multiplunge} as $N\to\infty$,
with an $O(1)$ per-edge offset that keeps the finite-$N$ ratio near-integer but not exactly
integer at the sizes shown.

Two limits bound this regime. As $M$ grows the bands fill the axis: the upper cluster fraction
$2WM$ approaches $1$ and the operator approaches the identity. The plunge stays a vanishing
fraction only while $M\log N$ is small compared with $N$, so a growing number of bands must also
narrow.

\begin{figure}[t]
\centering
\includegraphics[width=0.65\linewidth]{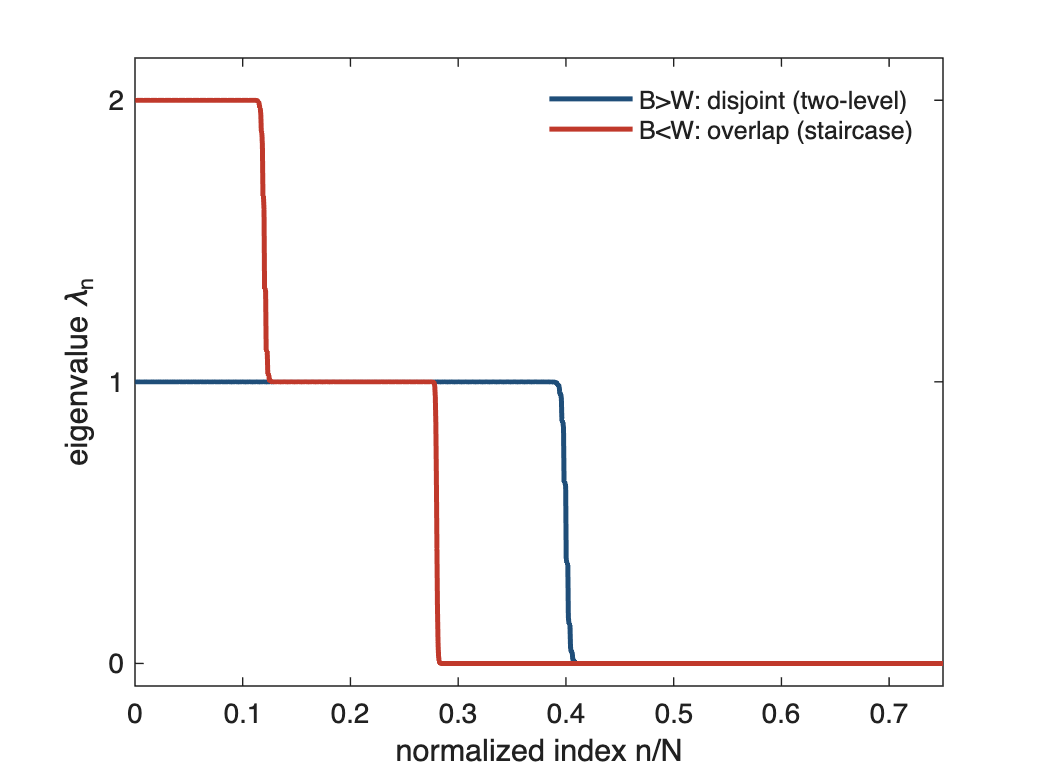}
\caption{Frequency pulse train of four equal pulses ($N=2000$). For $B>W$ the pulses are
disjoint and the spectrum is two-level; for $B<W$ they overlap, the symbol becomes
multilevel, and the single plunge breaks into a staircase (here a $\lambda\approx2$ step
on the double-overlap band).}
\label{fig:overlap}
\end{figure}

The transition from two clusters to \emph{staggered steps} is controlled by the
geometry of the train. With equal pulses of half-width $W$ and centers spaced by $2B$, both in the fractional units
of \eqref{eq:box},
adjacent pulses are disjoint for $B\ge W$ and overlap for $B<W$, their nearest edges being
separated by $2(B-W)$. Because the symbol is the
sum of the individual pulse symbols, on a frequency covered by several overlapping pulses the
symbol equals $h$ times their number: overlapping pass bands stack their heights. In the disjoint case
($B\ge W$) no frequency is covered twice, the symbol stays $\{0,h\}$-valued, and the
spectrum is two-level with a single plunge. When $B<W$ the pulses overlap, the symbol takes
the integer-multiple heights $\{0,h,2h,\dots\}$, and by the Szeg\H{o} reading the spectrum
acquires a cluster at every attained height, separated by a Landau--Widom plunge at each
riser, a staircase. Fig.~\ref{fig:overlap} contrasts the two cases for a four-pulse
train, the overlapping case showing a clear $\lambda\approx2$ plateau where two pulses
coincide. The opposite limit, $W\to0$, is the atomic regime; carrying the overlap idea to a
self-similar arrangement of heights gives a fractal staircase of clusters.

\subsection{Buried jumps: the pedestal plunge}
\label{sec:hat}
\begin{figure}[t]
\centering
\begin{subfigure}{0.49\linewidth}\includegraphics[width=\linewidth]{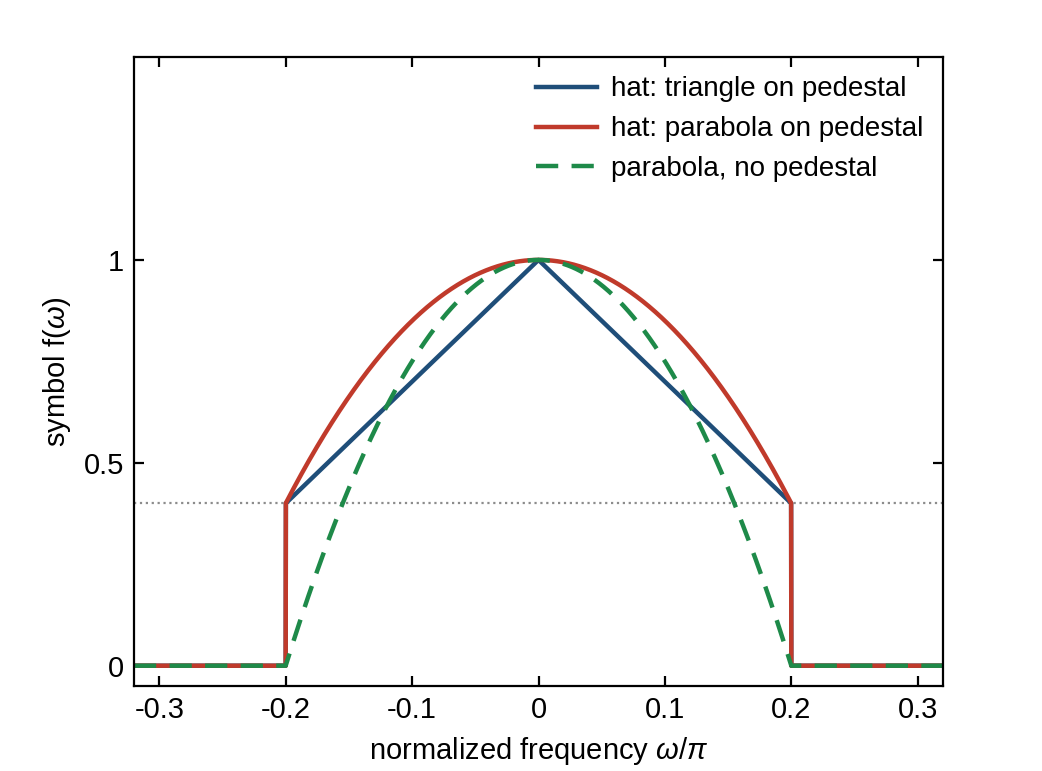}\caption{}\end{subfigure}\hfil
\begin{subfigure}{0.49\linewidth}\includegraphics[width=\linewidth]{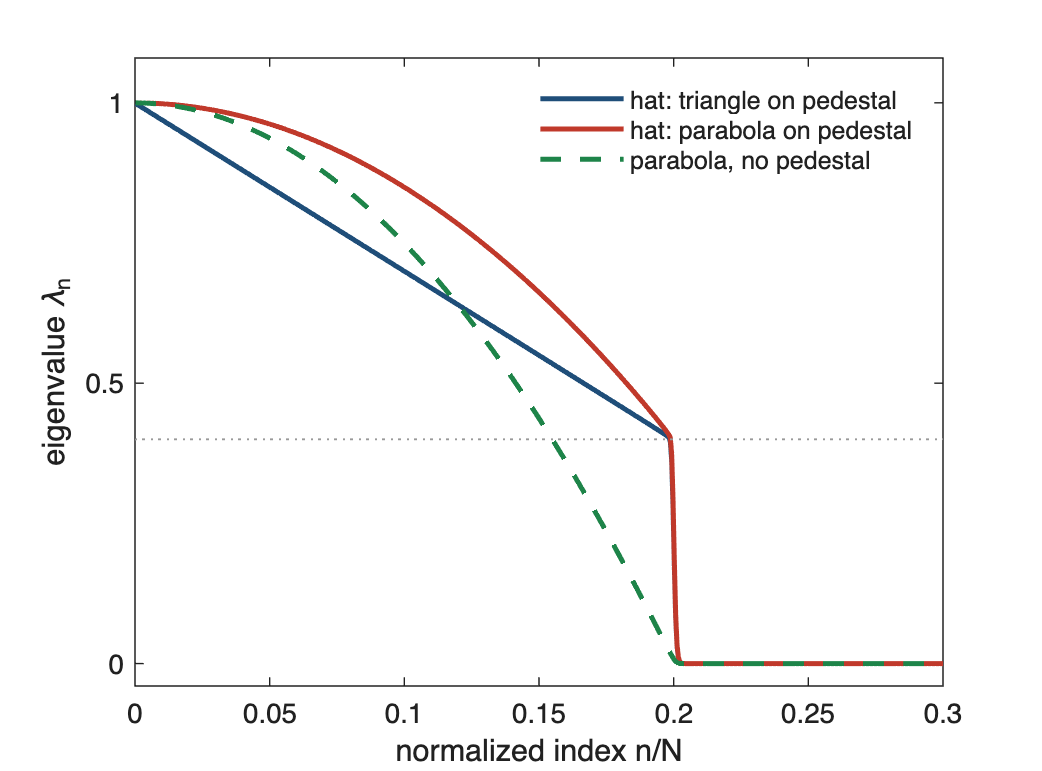}\caption{}\end{subfigure}
\caption{Hat symbols, pedestal height $h_0=0.4$, $N=2000$. (a) Frequency symbols: a
triangular and a parabolic cap on a rectangular pedestal (both jump $h_0\to0$ at the edge),
and a pedestal-free parabola that vanishes continuously at the edge. (b) Eigenvalue spectra:
both hats plunge at $\lambda=h_0$ regardless of cap shape, while the pedestal-free parabola
descends smoothly through $h_0$ with no plunge.}
\label{fig:hat}
\end{figure}

A jump need not sit at the top of the spectrum to make a plunge. Consider a ``hat''
symbol: a rectangular pedestal of height $h_0$ on $|\omega|<2\pi W$ carrying a smooth cap (a
triangle or a parabola) that rises to a peak at the band center. The cap is continuous and
contributes no jump, but the pedestal still drops discontinuously from $h_0$ to $0$ at the
band edge. By the per-edge mechanism that jump produces a Landau--Widom plunge, now located
not at the top of the spectrum but at the level $\lambda=h_0$, the pedestal height. Above
$h_0$ the smooth cap fills the spectrum continuously, exactly as in the smooth regime; the plunge is
the thin transition from $h_0$ down to the zero cluster (Fig.~\ref{fig:hat}).

The cap shape is irrelevant to the plunge. A triangular and a parabolic cap give the
identical plunge, consistent with a very thin logarithmic plunge at these sizes, and differ only
in how they fill the band above
$h_0$, where the eigenvalue count instead grows like $N$. Removing the pedestal, so that the
cap vanishes at the band edge, removes the jump and with it the plunge, leaving a smooth
descent through $h_0$ (the dashed curve in Fig.~\ref{fig:hat}). The lesson, which holds
across all the regimes above, is that a plunge appears at every jump of the symbol, sited at
the level of that jump and with width set by the jump alone, regardless of whatever smooth
structure is layered on top of it.

\section{Self-Similar Symbols and the Cantor Dimension}
\label{sec:devil}
Multilevel indicator symbols are governed entirely by the Szeg\H{o} push-forward: a symbol
that is a staircase, taking value $v_\ell$ on a frequency set of measure $\mu_\ell$,
produces eigenvalues that cluster at each height $v_\ell$ with cluster fraction $\mu_\ell$,
the clusters separated by Landau--Widom plunges at the risers. An arithmetic staircase
(nested pulses with equal height increments) gives equally spaced clusters; unequal
heights give a staggered set. The integrated density of states (IDOS) is then a step
function whose jumps sit at the attained heights.

The self-similar case is the interesting one, and it behaves in a way that at first looks
paradoxical. Let the frequency symbol be the Cantor--Lebesgue function (devil's staircase)
on the band, $f(\omega)=C(|\omega|/\pi)$, where $C$ is the middle-thirds
Cantor function. This $f$ is continuous, yet its limiting eigenvalue distribution is purely
atomic: a cluster ladder, not a smooth fill. The reason lies in how the Szeg\H{o}
push-forward weights the symbol's values. The Cantor function rises only on the Cantor set,
which has measure zero, and is flat on the complementary intervals, whose lengths sum to
one. So although $f$ is continuous, it spends essentially all of its ``frequency budget''
sitting on flat plateaus and only a measure-zero fraction climbing between them. By the
histogram reading of Szeg\H{o}, the eigenvalues inherit exactly this weighting: they pile up
at the plateau heights and almost none land on the connecting ramps. The push-forward of
Lebesgue measure under $C$ is therefore \emph{purely atomic}, with an atom at each
plateau value, a continuous symbol whose limiting eigenvalue distribution is purely atomic.

\begin{proposition}[Cantor-dimension cluster law]
\label{prop:devil}
For the devil's-staircase symbol the eigenvalues of $T_N(f)$ cluster at the dyadic plateau
heights $\{(2j-1)/2^\ell\}$. The cluster at a level-$\ell$ plateau has fractional mass
$m=3^{-\ell}$, and there are $2^{\ell-1}$ such clusters, so the number of clusters of mass
$m$ obeys
\begin{equation}
\#\{\text{clusters of mass }m\}=\tfrac12\,m^{-D},\qquad D=\frac{\log 2}{\log 3}=0.6309\ldots,
\label{eq:cantor}
\end{equation}
the Hausdorff dimension of the Cantor set \cite{falconer}. The limiting spectral measure is pure
point, and its integrated density of states is a self-similar pure-jump staircase, with a jump of
$3^{-\ell}$ at each level-$\ell$ plateau height.
\end{proposition}

\begin{proof}
By Szeg\H{o}, the spectral measure converges to $C_*(\text{Leb})$. Since $C$ is constant
($=v$) on each removed interval $I$ of length $|I|$, the preimage $C^{-1}(\{v\})\supseteq I$
has positive measure, so $v$ carries an atom of mass $\ge|I|$; as the removed intervals
have total length one, the limiting measure is purely atomic with masses equal to the
plateau lengths. In the middle-thirds construction the level-$\ell$ plateaus number
$2^{\ell-1}$ and each has length $3^{-\ell}$, giving \eqref{eq:cantor} on eliminating
$\ell$. The IDOS, which jumps by $m$ at each plateau height, inherits the self-similar
step structure of $C$.
\end{proof}

\begin{figure}[t]
\centering
\begin{subfigure}{0.32\linewidth}\includegraphics[width=\linewidth]{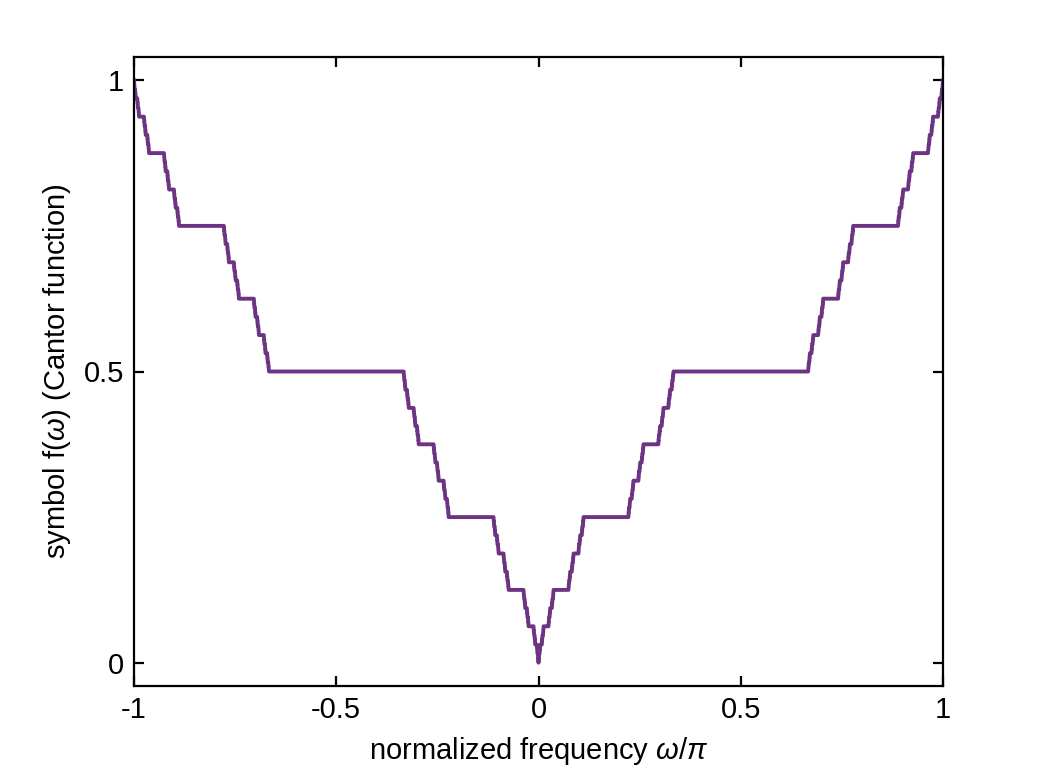}\caption{}\end{subfigure}\hfil
\begin{subfigure}{0.32\linewidth}\includegraphics[width=\linewidth]{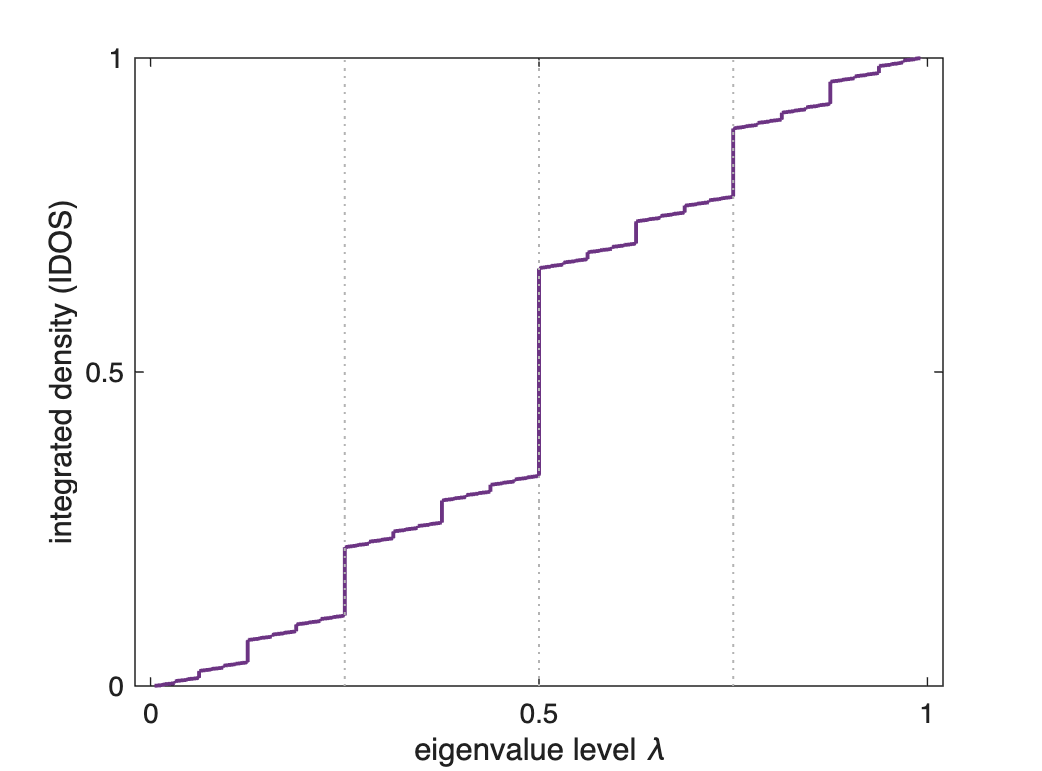}\caption{}\end{subfigure}\hfil
\begin{subfigure}{0.32\linewidth}\includegraphics[width=\linewidth]{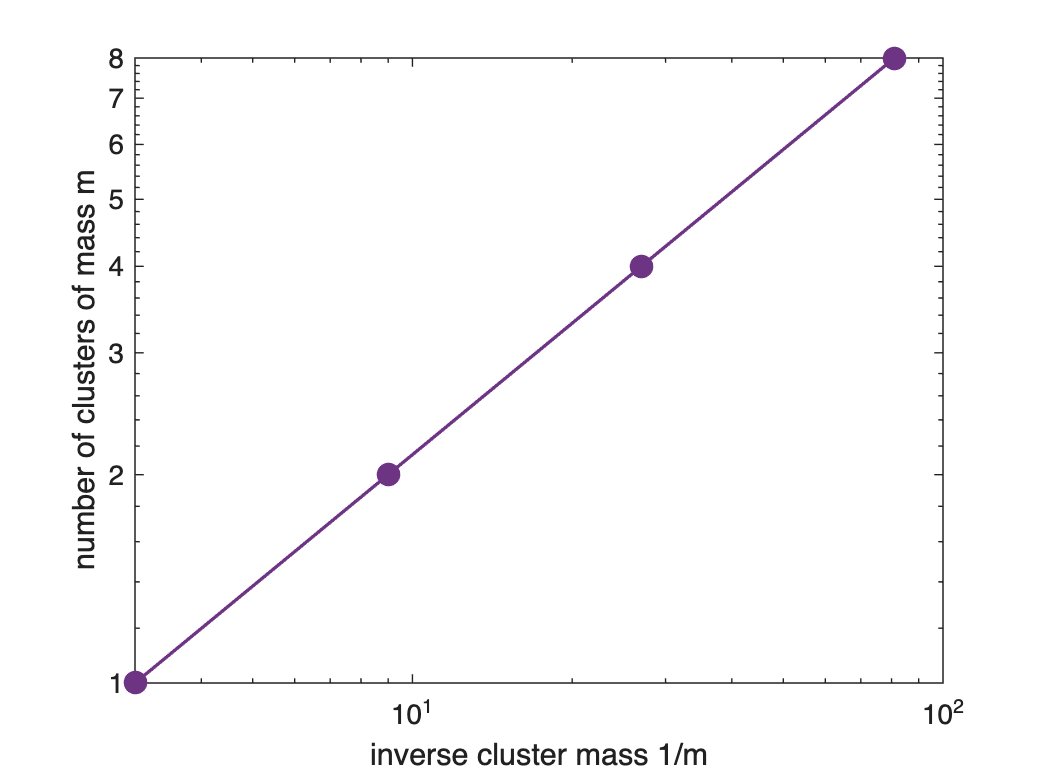}\caption{}\end{subfigure}
\caption{Devil's-staircase symbol, $N=2000$. (a) The Cantor-function symbol $f(\omega)$ in
the frequency domain. (b) The empirical integrated density of states of $T_N(f)$, a self-similar
pure-jump staircase, with the largest jumps at the dyadic levels $\tfrac14,\tfrac12,\tfrac34$.
(c) Number of eigenvalue clusters of mass $m$ versus $1/m$ at the Cantor masses $m=3^{-\ell}$,
on log--log axes; the slope is the Cantor dimension $D=\log2/\log3$.}
\label{fig:devil}
\end{figure}

Fig.~\ref{fig:devil} confirms this at $N=2000$. The measured cluster masses in
Fig.~\ref{fig:devil}(b) agree with the Cantor values $3^{-\ell}$ to within the finite-$N$
resolution of the clusters, and the cluster counts per level are $2^{\ell-1}$.
Fig.~\ref{fig:devil}(c) plots the resulting count-versus-mass law at the Cantor masses. The exponent is a dimension in the literal sense: tallying
clusters by how small their mass is is the same scale-by-scale counting that defines the
Cantor set's fractal dimension, so the eigenvalue bookkeeping and the geometry of the
symbol's support are governed by one and the same number. Unlike the kernel-dependent
prefactor $C_p$ of Theorem~\ref{thm:edge} or the integer $2M$ of the atomic regime, $D$ is
the Hausdorff dimension of the Cantor set underlying the symbol and is independent of the vertical
labeling of the plateau heights in this Cantor construction. Proposition~\ref{prop:devil} follows directly from the
Szeg\H{o} push-forward of Section~\ref{sec:setup} together with the classical fact that the
push-forward of Lebesgue measure under the Cantor function is purely atomic; the taxonomy
thus reaches a fractal dimension in the concentration-operator setting.

\section{Atomic Symbols and Finite Rank}
\label{sec:atomic}
When each band collapses to a spectral line the symbol becomes atomic and the operator
becomes finite rank: there is no cluster at the top (a point mass carries zero measure) and
no plunge. The rank statement is classical: it is Carath\'eodory's finite-rank decomposition of a
positive semidefinite Toeplitz form \cite{grenanderszego}, the basis of Pisarenko's harmonic
decomposition \cite{pisarenko73}. The closed-form eigenvalues follow from a $2\times2$ Gram matrix.

\begin{proposition}[Cosine kernels are finite rank]
\label{prop:cos}
The Toeplitz matrix $T_N(r)$ with $r_k=\cos(\omega_0 k)$ and $0<\omega_0<\pi$ has rank $2$, with nonzero
eigenvalues
\begin{equation}
\lambda_\pm=\frac{N}{2}\pm\frac12\left|\frac{\sin(N\omega_0)}{\sin\omega_0}\right| ,
\label{eq:cos}
\end{equation}
and $N-2$ zero eigenvalues. A sum of $M$ cosines at distinct frequencies in $(0,\pi)$ has rank $2M$ once $N\ge 2M$.
\end{proposition}

\begin{proof}
Writing $u_m=\cos(\omega_0 m)$, $v_m=\sin(\omega_0 m)$ one has
$[T_N(r)]_{mn}=\cos(\omega_0(m-n))=u_mu_n+v_mv_n$, so $T_N(r)=\mathbf u\mathbf u^{\!\top}
+\mathbf v\mathbf v^{\!\top}$ has rank at most $2$. Its nonzero eigenvalues are those of
the $2\times2$ Gram matrix of $\{\mathbf u,\mathbf v\}$, whose trace is $N$ and whose
eigenvalues are $\tfrac N2\pm\tfrac12|\sum_{m=0}^{N-1}e^{2i\omega_0 m}|$, giving
\eqref{eq:cos} through the Dirichlet kernel. For $M$ cosines the columns $e^{\pm i\omega_j m}$,
$j=1,\dots,M$, form a Vandermonde family with $2M$ distinct nodes, so for $N\ge 2M$ the Gram
matrix has rank $2M$.
\end{proof}

\begin{figure}[t]
\centering
\includegraphics[width=0.6\linewidth]{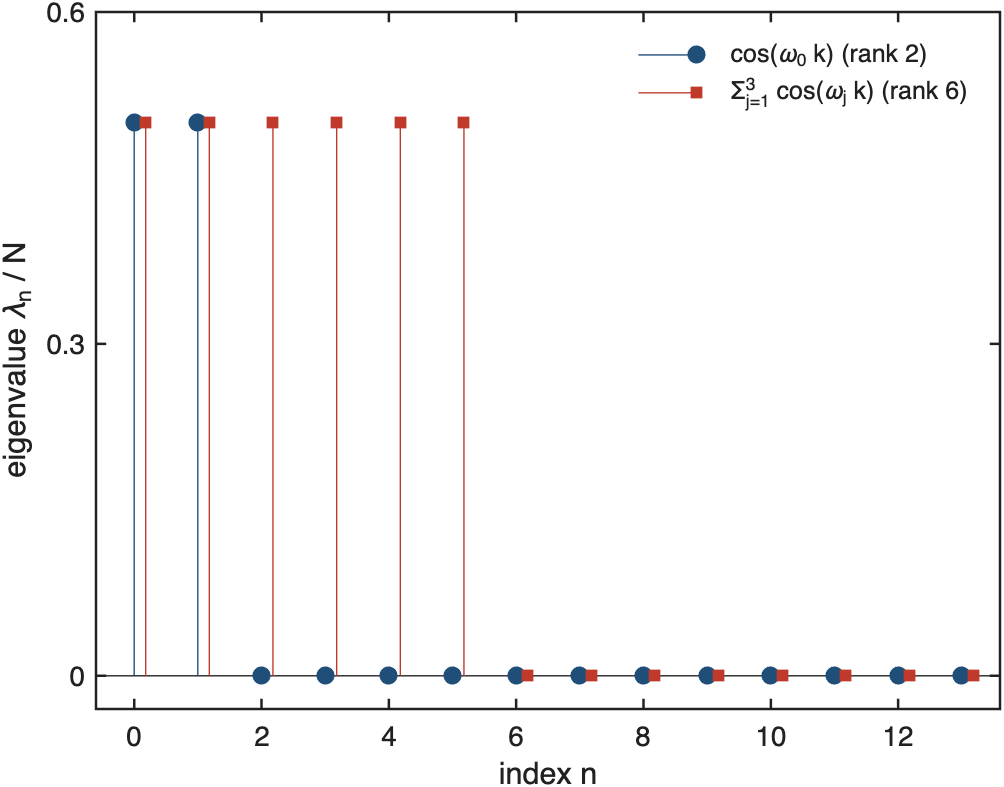}
\caption{Atomic regime, $N=2000$. Normalized eigenvalues of cosine-sum kernels: $\cos(\omega_0 k)$
has rank $2$ and $\sum_{j=1}^3\cos(\omega_j k)$ has rank $6$. Exactly $2M$ eigenvalues sit at
$\approx N/2$; the remainder are zero, a rank cliff, not a plunge.}
\label{fig:atomic}
\end{figure}

This is the zero-width limit of the indicator regime: a band of width $2W$ carries $\approx2WN$
eigenvalues near $1$; as $W\to0$ that block shrinks until, in the strict limit, each
symmetric line pair contributes a single rank-$2$ piece. Fig.~\ref{fig:atomic} shows the
rank cliff for $M=1$ and $M=3$: exactly $2M$ eigenvalues at $\approx N/2$, the rest
identically zero, with no graded transition.

\begin{remark}[Plunge versus cliff]
The atomic regime sharpens the distinction between the two ways a spectrum can fall from
large to small. A \emph{plunge} is soft: an $O(\log N)$ ramp of eigenvalues sits strictly
between the clusters (a number that grows with $N$), the operator is of full rank $N$ and
is only an \emph{approximate} projection, and the transition is the finite-$N$ image of a
jump in the symbol. A \emph{cliff} is hard: exactly $N-2M$ eigenvalues are identically zero
with none in between for any $N$, the operator is \emph{exactly} of rank $2M$, and the
symbol is atomic. The cliff is not a narrow plunge but the singular endpoint $W\to0$: for
every positive bandwidth the box retains a $\log N$ plunge (whose count is independent of
$W$), and only at zero width does it collapse to the exact-rank cliff.
\end{remark}

\begin{remark}[Subspace methods]
The rank-$2M$ structure is the algebraic basis of line-spectrum and subspace estimation. $M$
well-separated sinusoids span a signal subspace of dimension $2M$, and the absence of a plunge
affords a clean split of signal from noise for resolvable tones.
\end{remark}

\section{Eigenvector Zeros for the Single Band}
\label{sec:zeros}
The eigenvalues fix the value distribution of the symbol. The eigenvectors carry a further
classical fact. Order the eigenvalues $\lambda_0\ge\cdots\ge\lambda_{N-1}$ with eigenvectors
$v^{(k)}$, and form the eigenfilter polynomial $V_k(z)=\sum_{n=0}^{N-1}v^{(k)}_n z^{n}$ of
degree $N-1$.

By Trench's Theorem~8, reached through his Theorem~6, a Hermitian Toeplitz matrix whose symbol is
real, even, and monotonic on $(0,\pi)$ has simple eigenvalues, and every eigenfilter $V_k$ has all
$N-1$ of its zeros on the unit circle $|z|=1$ \cite{trench94}. The single band and every B-spline
power $f_p$ of Section~\ref{sec:smooth} satisfy this hypothesis, so their eigenfilter zeros are all
on the circle. For the single band, Slepian's discrete prolate analysis \cite{slepian78} fixes how
those on-circle zeros split between passband and stopband: the $k$-th eigenfilter carries $k$ zeros
inside the band. Proposition~\ref{prop:zeros} combines the two classical facts, Slepian's count and
Trench's placement, and records the conjugate-pair bookkeeping and the lowpass reading that follow.

\begin{proposition}[Passband share of the single-band eigenfilters]
\label{prop:zeros}
For the single-band prolate matrix $T_N(a)$, order the eigenvalues by decreasing concentration.
The $k$-th eigenfilter $V_k$ has exactly $k$ of its $N-1$ on-circle zeros in the passband
$|\omega|<2\pi W$ and the rest in the stopband: $\lfloor k/2\rfloor$ conjugate pairs, and for odd
$k$ one further zero at $z=1$ ($\omega=0$, the band center), which the antisymmetry of the odd
orders forces for every $N$. The even orders are symmetric and the odd orders antisymmetric. The
only other forced zero is a stopband zero at $z=-1$ ($\omega=\pi$), carried by an odd order when
$N$ is odd and by an even order when $N$ is even.
\end{proposition}

\begin{proof}
By Trench's Theorem~8 the $N-1$ zeros lie on the unit circle \cite{trench94}. The symbol is real
and even, so $T_N(a)$ is centrosymmetric; it commutes with Slepian's tridiagonal (Jacobi) matrix
\cite{slepian78}, which has positive off-diagonals and a simple spectrum, so by Sturm
oscillation the $k$-th eigenvector has exactly $k$ sign changes; centrosymmetry then gives it
parity $(-1)^k$, symmetric for even $k$ and antisymmetric for odd $k$. On the unit circle
the linear-phase amplitude of $V_k$ is Slepian's discrete prolate spheroidal wave function, which
has exactly $k$ zeros in the open passband $(-2\pi W,2\pi W)$ \cite{slepian78}.
Parity makes these passband zeros symmetric about $\omega=0$, with one forced at $\omega=0$ (that
is, $z=1$) when $k$ is odd. The $k$ passband zeros group into $\lfloor k/2\rfloor$ conjugate pairs,
the central zero at $z=1$ left unpaired for odd $k$.
\end{proof}

\begin{figure}[t]
\centering
\begin{subfigure}{0.47\linewidth}\includegraphics[width=\linewidth]{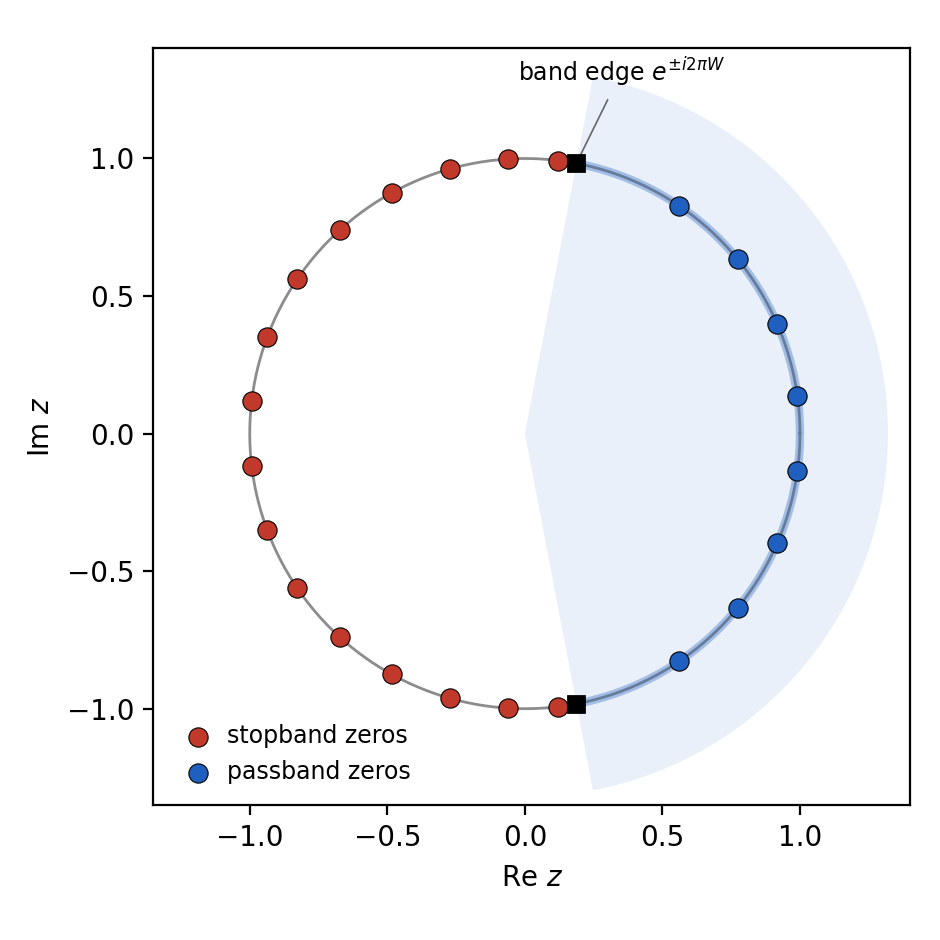}\caption{}\end{subfigure}\hfil
\begin{subfigure}{0.49\linewidth}\includegraphics[width=\linewidth]{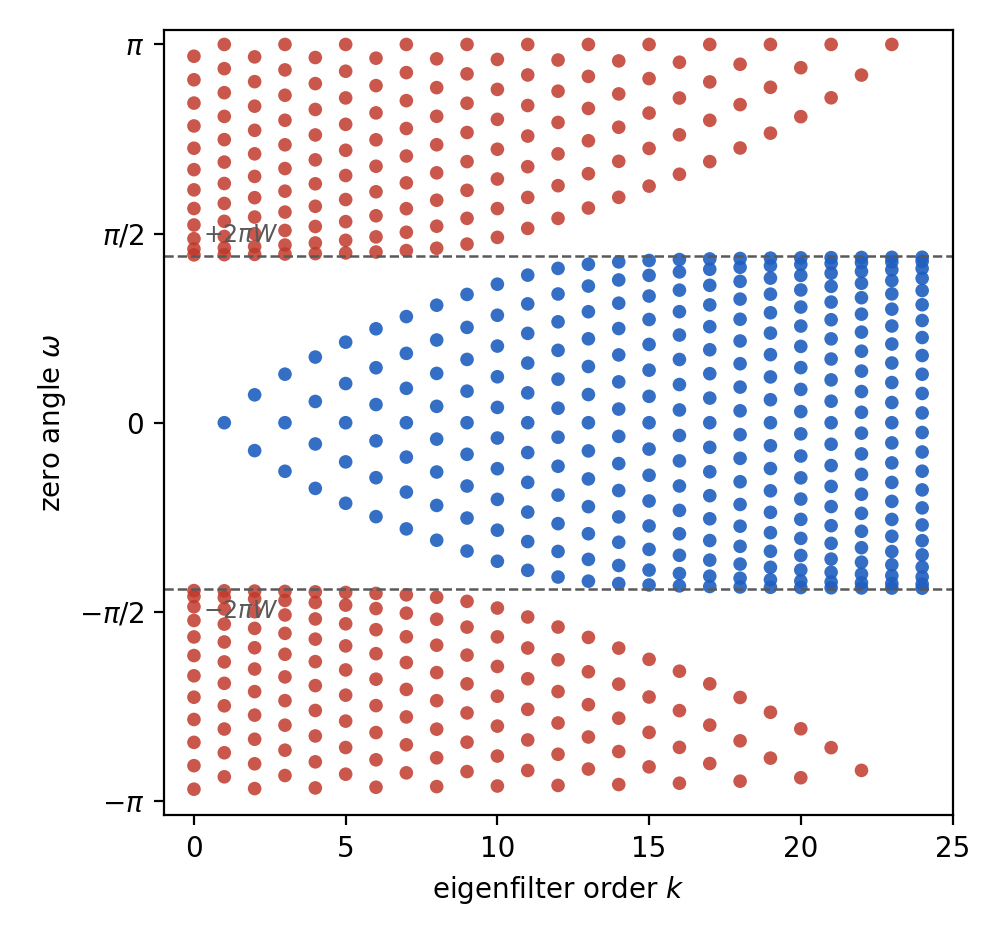}\caption{}\end{subfigure}
\caption{On-circle eigenfilter zeros split by band, single band ($N=25$, $W=0.22$), zeros computed
from the commuting Jacobi matrix. (a) The $k=8$ eigenfilter: of its on-circle zeros,
$\lfloor k/2\rfloor$ conjugate pairs fall in the passband $|\omega|<2\pi W$ (blue, shaded), the
rest in the stopband (red). (b) Zero angle against eigenfilter order: the passband zeros (blue)
grow as $\lfloor k/2\rfloor$ pairs while the stopband zeros (red) recede, so low-order eigenfilters
are lowpass and high-order ones highpass. All zeros lie on the circle (Trench).}
\label{fig:zeros}
\end{figure}

Read the entries of an eigenvector as filter taps and factor the resulting polynomial: all its
roots lie on the unit circle \cite{trench94}, and $k$ of them, that is $\lfloor k/2\rfloor$
conjugate pairs, fall inside the signal band \cite{slepian78}. The
dominant eigenvectors place their zeros in the stopband and act as lowpass filters. This is the
structural basis of root-MUSIC, where the on-circle zeros of the noise-subspace eigenfilters
mark the signal frequencies. Fig.~\ref{fig:zeros}(a) shows these on-circle zeros for one order, and
Fig.~\ref{fig:zeros}(b) the passband share $\lfloor k/2\rfloor$ against eigenfilter order.

\section{Discussion: Symbol Regularity and the Spectral Descriptor}
\label{sec:discussion}
The regimes are ordered by the regularity of the symbol, and the descriptor
of the spectral edge changes accordingly:
\begin{itemize}
\item \emph{Indicator (jump).} Two clusters, plunge of width $\propto\log N$, universal
\emph{constant} $1/\pi^2$ per band edge.
\item \emph{Smooth (B-spline of order $p$).} No plunge, algebraic edge with universal
\emph{exponent} $p-1$.
\item \emph{Atomic (point masses).} Finite rank, the universal descriptor being the
\emph{integer} $2M$.
\item \emph{Self-similar (devil's staircase).} A cluster hierarchy whose count-versus-mass
scaling is the \emph{dimension} $D=\log2/\log3$, the Hausdorff dimension of the underlying
Cantor set.
\end{itemize}
Landau--Widom serves as the common reference point for the indicator and smooth regimes at one
band ($M=1$) and unit power ($p=1$); the atomic cliff is a separate zero-width singular endpoint.

It is worth stating what the cluster-and-plunge structure means operationally. When the
symbol is a $\{0,1\}$ indicator the operator is asymptotically an orthogonal projection:
about $2WN$ eigenvalues sit at $1$ and the rest at $0$, so the matrix is effectively low
rank (of numerical rank $2WN$) even though it is formally of full rank $N$. The top
eigenvectors, the Slepian sequences whose eigenvalues cluster at $1$, span the effective
signal subspace and capture essentially all of the energy of signals that are simultaneously
time- and band-limited. The plunge is precisely the thin, $O(\log N)$ set of basis functions
that straddle the boundary, belonging neither fully inside nor fully outside the subspace;
its vanishing fraction is exactly why the low-rank picture is accurate. This is an
\emph{approximate} projection, in contrast to the atomic regime, where the operator is
\emph{exactly} finite rank $2M$: there the top subspace captures everything with no
remainder, and the absence of a plunge is the rank cliff marking an exact rather than
numerical dimension.

Multilevel plunges extend the same reading. A staircase symbol taking distinct heights
$v_\ell$ makes the operator asymptotically a weighted sum of mutually orthogonal
projections, $T_N\approx\sum_\ell v_\ell P_\ell$, where $P_\ell$ projects onto the cluster
at height $v_\ell$; each riser is the thin boundary between two consecutive cluster
subspaces. The operator thus acts as a graded bank of gains (one subspace per distinct
symbol level, the size of cluster $\ell$ being the measure of the level set $\{f=v_\ell\}$
times $N$), and the devil's staircase is the limiting case of countably many such levels
organized self-similarly. In every case the clustered subspaces carry the operator and the
plunges merely mark their boundaries; only when the levels collapse to point masses does the
structure become exactly finite rank.

Structurally, the absence of a $1/\pi^2$-type constant in the smooth regime is expected
rather than a gap: the logarithmic term that carries the Landau--Widom constant is a
Fisher--Hartwig effect of a jump, and for a continuous symbol the strong
Szeg\H{o}/Wiener--Hopf trace expansion has only a volume term and an $O(1)$ correction,
with no $\log N$ term to carry a universal coefficient \cite{charlierclaeys15,sobolev13}.
The smooth regime is thus characterized by the exponent $p-1$ alone.

\begin{figure}[t]
\centering
\includegraphics[width=0.65\linewidth]{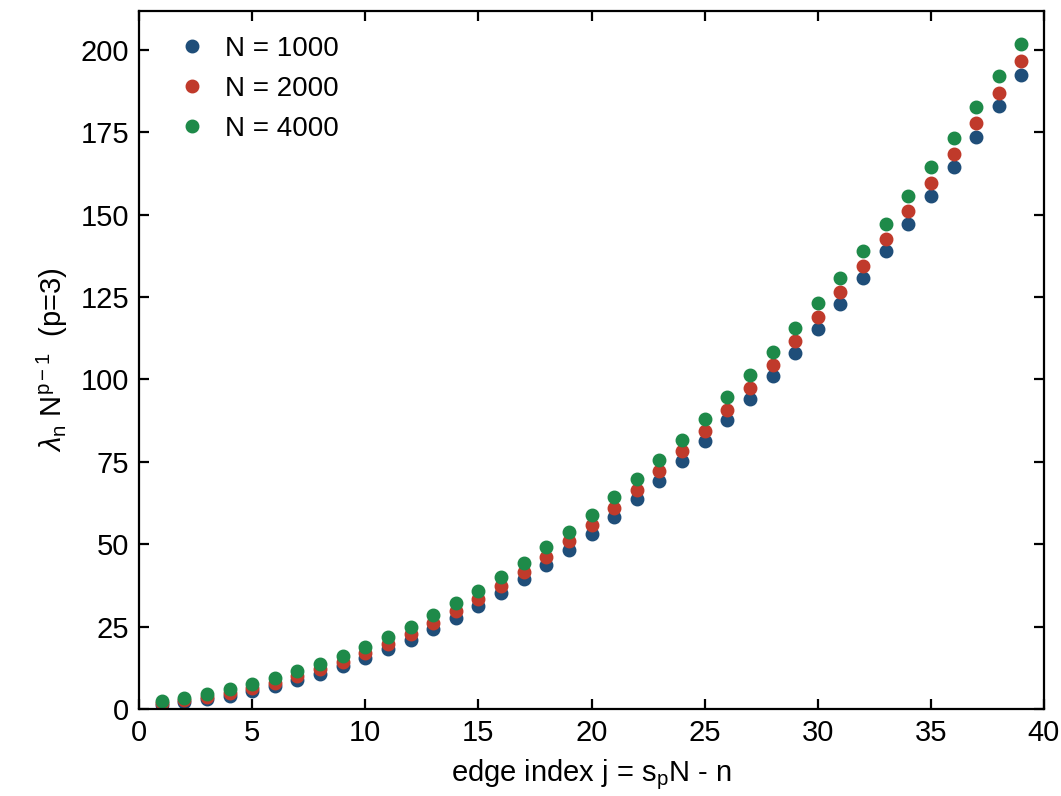}
\caption{Test of a universal edge collapse for the smooth ($p=3$) B-spline spectral edge, $W=0.05$. The
rescaled edge eigenvalues $\lambda_n N^{p-1}$ are plotted against the edge index $j=s_pN-n$ for
$N=1000,2000,4000$. The three curves do not collapse onto one: at fixed $j$ the rescaled values
grow with $N$.}
\label{fig:edgenocollapse}
\end{figure}

It is natural to ask whether a $1/\pi^2$-type universal constant also lives at the smooth
spectral edge. The simplest test carries the edge law to a fixed index $j=s_pN-n$ from the edge:
if \eqref{eq:edge} held there, the rescaled eigenvalues $\lambda_n N^{p-1}$ would depend on $j$
alone and the curves for different $N$ would collapse onto one. They do not.
Fig.~\ref{fig:edgenocollapse} shows that for $p=3$ the rescaled values grow with $N$ at fixed $j$
rather than converging, so the law of Theorem~\ref{thm:edge} does not extend to a fixed index.
The spectrum is deterministic rather than a random ensemble, so the random-matrix
Airy/Tracy--Widom mechanism \cite{tracywidom94} would not be expected to apply; the
microscopic edge appears instead to be organized by deterministic Toeplitz
quantization corrections \cite{bogoya15}, which carry $N$-dependent drifts. The dimensionless
constant in this landscape thus appears not at the smooth edge but at the
self-similar symbol, namely the Cantor dimension $D=\log2/\log3$ of
Proposition~\ref{prop:devil}.

\section{Conclusion}
The starting point is classical. A single rectangular band, observed through a finite window,
gives the two-cluster spectrum and the Landau--Widom plunge, whose universal coefficient is
$1/\pi^2$. This paper deforms that single box and follows the plunge. Smoothing the band edge by
Hadamard powers replaces the plunge with an algebraic edge of exponent $p-1$. Replicating the
band multiplies the plunge edge for edge, as the multiple-interval theorem of Landau and Widom
implies, so $1/\pi^2$ is a per-edge constant. Collapsing the band to a line yields the classical
finite-rank limit. The descriptor of the spectral edge passes from a constant to
an exponent to an integer as the symbol goes from discontinuous to smooth to atomic, with the
single box as the shared limiting case. Within the indicator class, overlapping the pulses ($B<W$) breaks the single
plunge into a multilevel staircase, and a self-similar symbol promotes the descriptor to a
fractal dimension, the Cantor dimension $D=\log2/\log3$ of the eigenvalue cluster
hierarchy. At the smooth edge, the eigenvalues rescaled by a power of $N$ do not collapse at a fixed index
from the edge, so no universal constant appears there. On the eigenvector side, two classical facts
combine for the single band: Slepian's zero count \cite{slepian78} places $k$ zeros of the $k$-th
eigenfilter inside the band, and Trench's theorem \cite{trench94} puts every eigenfilter zero on the
unit circle, so the dominant eigenvectors act as lowpass filters.

These regimes are not merely a taxonomy. They read as statements about the independent
signaling dimensions available in a fixed time--bandwidth budget. The eigenvalues of the
concentration operator count how many orthogonal waveforms can be simultaneously time- and
band-limited, the classical $\approx2WN$ degrees of freedom of Slepian's theory, and the
regularity of the symbol sets how that count behaves. A smooth edge replaces the sharp plunge
by a graceful roll-off, and its exponent measures how gently the marginal modes fade. Disjoint
bands multiply the ambiguous transition by the number of band edges while leaving the usable
cluster fixed. A line spectrum collapses the operator to exact finite rank. The four edge
descriptors, a constant, an exponent, an integer, and a fractal dimension, form a compact
vocabulary for how structured spectral occupancy sets the usable dimensions. The same structure
underlies pulse shaping, fragmented-spectrum and multiband systems, and the subspace methods of
line-spectrum estimation.

\appendix
\section{B-spline Endpoint Coefficient}
\label{app:bspline}
The symbol of $a^{\circ p}$ is the $p$-fold circular self-convolution of the box
$f_1=\mathbf 1_{|\omega|<2\pi W}$; equivalently $f_p=(2\pi)^{-(p-1)}\,g^{*p}$, where $g^{*p}$
is the ordinary $p$-fold convolution of the box $g=\mathbf 1_{[0,\ell]}$ with $\ell=4\pi W$. On its
outermost piece $0\le x\le \ell$ one has the exact monomial $g^{*p}(x)=x^{p-1}/(p-1)!$. Writing
$\epsilon=\omega_e-\omega$ for the distance inside the support endpoint $\omega_e=2\pi pW$,
the peak-normalized symbol is therefore
\[
\hat f_p(\omega)=\frac{f_p(\omega)}{f_p(0)}=\frac{g^{*p}(\epsilon)}{g^{*p}(\mathrm{center})}
=C_p\,\epsilon^{\,p-1},
\]
\[
C_p=\frac{1}{(p-1)!\,g^{*p}(\mathrm{center})},
\]
exact on $0\le\epsilon\le \ell$, the factor $(2\pi)^{-(p-1)}$ cancelling in the ratio.
Equivalently, in terms of the symbol peak $f_p(0)=\sum_k a_k^p$ (asymptotically the largest
eigenvalue),
\[
C_p=\frac{1}{(p-1)!\,(2\pi)^{p-1}\,f_p(0)} ,
\]
since $g^{*p}(\mathrm{center})=(2\pi)^{p-1}f_p(0)$. For example $g^{*2}(\mathrm{center})=\ell$,
$g^{*3}(\mathrm{center})=\tfrac34 \ell^2$, $g^{*4}(\mathrm{center})=\tfrac23 \ell^3$, giving
$C_2=1/\ell$, $C_3=2/(3\ell^2)$, $C_4=1/(4\ell^3)$, which match the fitted edge constants. Only the
exponent $p-1$ is universal; $C_p$ depends on $p$, on $W$, and on the chosen normalization.

\section*{AI Acknowledgment}
AI tools were used mainly for editorial assistance including proofreading, literature search and
language refinement. All theoretical developments, derivations and conclusions are the author's work.

\section*{Data Availability}
The data that support the findings of this study are available from the author upon reasonable
request.


\relax


\begin{thebibliography}{99}
\bibitem{slepian61} D. Slepian and H. O. Pollak, ``Prolate spheroidal wave functions,
Fourier analysis and uncertainty---I,'' \emph{Bell Syst. Tech. J.}, vol.~40, no.~1,
pp.~43--63, 1961.
\bibitem{landaupollak61} H. J. Landau and H. O. Pollak, ``Prolate spheroidal wave
functions, Fourier analysis and uncertainty---II,'' \emph{Bell Syst. Tech. J.}, vol.~40,
no.~1, pp.~65--84, 1961.
\bibitem{landaupollak62} H. J. Landau and H. O. Pollak, ``Prolate spheroidal wave
functions, Fourier analysis and uncertainty---III,'' \emph{Bell Syst. Tech. J.}, vol.~41,
no.~4, pp.~1295--1336, 1962.
\bibitem{slepian78} D. Slepian, ``Prolate spheroidal wave functions, Fourier analysis and
uncertainty---V: The discrete case,'' \emph{Bell Syst. Tech. J.}, vol.~57, no.~5,
pp.~1371--1430, 1978.
\bibitem{landauwidom80} H. J. Landau and H. Widom, ``Eigenvalue distribution of time and
frequency limiting,'' \emph{J. Math. Anal. Appl.}, vol.~77, no.~2, pp.~469--481, 1980.
\bibitem{widom82} H. Widom, ``On a class of integral operators with discontinuous symbol,''
in \emph{Toeplitz Centennial}, Oper. Theory Adv. Appl., vol.~4, pp.~477--500. Basel:
Birkh\"auser, 1982.
\bibitem{sobolev13} A. V. Sobolev, ``Pseudo-differential operators with discontinuous
symbols: Widom's conjecture,'' \emph{Mem. Amer. Math. Soc.}, vol.~222, no.~1043, 2013.
\bibitem{gioevklich06} D. Gioev and I. Klich, ``Entanglement entropy of fermions in any
dimension and the Widom conjecture,'' \emph{Phys. Rev. Lett.}, vol.~96, no.~10, 100503,
2006.
\bibitem{kms53} M. Kac, W. L. Murdock, and G. Szeg\H{o}, ``On the eigenvalues of certain Hermitian
forms,'' \emph{J. Rational Mech. Anal.}, vol.~2, pp.~767--800, 1953.
\bibitem{grenanderszego} U. Grenander and G. Szeg\H{o}, \emph{Toeplitz Forms and Their
Applications}. Berkeley, CA: Univ. California Press, 1958.
\bibitem{pisarenko73} V. F. Pisarenko, ``The retrieval of harmonics from a covariance function,''
\emph{Geophys. J. R. Astron. Soc.}, vol.~33, no.~3, pp.~347--366, 1973.
\bibitem{trench94} W. F. Trench, ``Some spectral properties of Hermitian Toeplitz matrices,''
\emph{SIAM J. Matrix Anal. Appl.}, vol. 15, no. 3, pp. 938--942, 1994.
\bibitem{papoulispillai} A. Papoulis and S. U. Pillai, \emph{Probability, Random Variables
and Stochastic Processes}, 4th ed. New York: McGraw-Hill, 2002.
\bibitem{osipov13} A. Osipov, V. Rokhlin, and H. Xiao, \emph{Prolate Spheroidal Wave
Functions of Order Zero}. New York: Springer, 2013.
\bibitem{davenport21} S. Karnik, J. Romberg, and M. A. Davenport, ``Improved bounds for
the eigenvalues of prolate spheroidal wave functions and discrete prolate spheroidal
sequences,'' \emph{Appl. Comput. Harmon. Anal.}, vol.~55, pp.~97--128, 2021.
\bibitem{bonami21} A. Bonami, P. Jaming, and A. Karoui, ``Non-asymptotic behavior of the
spectrum of the sinc kernel operator and related applications,'' \emph{J. Math. Phys.},
vol.~62, no.~3, 033511, 2021.
\bibitem{schoenberg73} I. J. Schoenberg, \emph{Cardinal Spline Interpolation}.
Philadelphia, PA: SIAM, 1973.
\bibitem{deboor01} C. de Boor, \emph{A Practical Guide to Splines}, rev.\ ed. New York:
Springer, 2001.
\bibitem{bottchersilbermann} A. B\"ottcher and B. Silbermann, \emph{Introduction to Large
Truncated Toeplitz Matrices}. New York: Springer, 1999.
\bibitem{gray06} R. M. Gray, ``Toeplitz and circulant matrices: A review,'' \emph{Found.
Trends Commun. Inf. Theory}, vol.~2, no.~3, pp.~155--239, 2006.
\bibitem{parter61} S. V. Parter, ``Extreme eigenvalues of Toeplitz forms and applications to
elliptic difference equations,'' \emph{Trans. Amer. Math. Soc.}, vol.~99, pp.~153--192, 1961.
\bibitem{widom61} H. Widom, ``Extreme eigenvalues of translation kernels,'' \emph{Trans. Amer.
Math. Soc.}, vol.~100, no.~2, pp.~252--262, 1961.
\bibitem{bottchergrudsky98} A. B\"ottcher and S. M. Grudsky, ``On the condition numbers of large
semi-definite Toeplitz matrices,'' \emph{Linear Algebra Appl.}, vol.~279, pp.~285--301, 1998.
\bibitem{garoni14} C. Garoni, C. Manni, F. Pelosi, S. Serra-Capizzano, and H. Speleers, ``On the
spectrum of stiffness matrices arising from isogeometric analysis,'' \emph{Numer. Math.},
vol.~127, no.~4, pp.~751--799, 2014.
\bibitem{proakis} J. G. Proakis and M. Salehi, \emph{Digital Communications}, 5th ed.
New York: McGraw-Hill, 2008.
\bibitem{falconer} K. Falconer, \emph{Fractal Geometry: Mathematical Foundations and
Applications}, 3rd ed. Chichester, U.K.: Wiley, 2014.
\bibitem{charlierclaeys15} C. Charlier and T. Claeys, ``Asymptotics for Toeplitz
determinants: Perturbation of symbols with a gap,'' \emph{J. Math. Phys.}, vol.~56, no.~2,
022705, 2015.
\bibitem{tracywidom94} C. A. Tracy and H. Widom, ``Level-spacing distributions and the
Airy kernel,'' \emph{Commun. Math. Phys.}, vol.~159, no.~1, pp.~151--174, 1994.
\bibitem{bogoya15} J. M. Bogoya, A. B\"ottcher, S. M. Grudsky, and E. A. Maximenko,
``Eigenvalues of Hermitian Toeplitz matrices with smooth simple-loop symbols,'' \emph{J.
Math. Anal. Appl.}, vol.~422, no.~2, pp.~1308--1334, 2015.
\end{thebibliography}
\end{document}